\documentclass[journal,comsoc]{IEEEtran}
\usepackage{soul}
\usepackage[T1]{fontenc}
\usepackage{cite}
\usepackage{amsmath, nccmath}
\usepackage{amsthm}
\usepackage[cmintegrals]{newtxmath}
\usepackage{algorithm}
\usepackage{array}
\usepackage{stfloats}
\usepackage{url}
\usepackage{comment}
\usepackage{hyperref}
\usepackage{cleveref}
\usepackage{multicol}
\usepackage{authblk}
\usepackage{graphicx}
\usepackage{caption}
\usepackage{textcomp}
\usepackage{xcolor}
\usepackage[normalem]{ulem}
\useunder{\uline}{\ul}{}
\usepackage{listings}
\usepackage{subcaption}
\usepackage{dirtytalk}
\usepackage{booktabs}
\usepackage{pifont}
\usepackage{multirow}
\usepackage{adjustbox}
\usepackage{pdflscape}
\usepackage{rotating}
\usepackage[utf8]{inputenc}
\usepackage{algpseudocode}
\usepackage{longtable}
\usepackage{enumitem}
\usepackage{tabularx}
\usepackage{xurl}
\def\BibTeX{{\rm B\kern-.05em{\sc i\kern-.025em b}\kern-.08em
    T\kern-.1667em\lower.7ex\hbox{E}\kern-.125emX}}

\usepackage[acronyms,nonumberlist,nopostdot,nomain,nogroupskip,acronymlists={hidden}]{glossaries}
\newglossary[algh]{hidden}{acrh}{acnh}{Hidden Acronyms}
\glsdisablehyper

\usepackage{tikz}
\usepackage{pgfplots}
\usepackage{glossaries}

\pgfplotsset{compat=newest}
\pgfplotsset{plot coordinates/math parser=false}
\newlength\fheight
\newlength\fwidth
\usetikzlibrary{plotmarks,patterns,decorations.pathreplacing,backgrounds,calc,arrows,arrows.meta,spy,matrix,scopes}
\usepgfplotslibrary{patchplots,groupplots}
\usepackage{tikzscale}

\newacronym{3gpp}{3GPP}{3rd Generation Partnership Project}
\newacronym{4g}{4G}{4th generation}
\newacronym{5g}{5G}{5th generation}
\newacronym{5gc}{5GC}{5G Core}
\newacronym{adc}{ADC}{Analog to Digital Converter}
\newacronym{aerpaw}{AERPAW}{Aerial Experimentation and Research Platform for Advanced Wireless}
\newacronym{ai}{AI}{Artificial Intelligence}
\newacronym{aimd}{AIMD}{Additive Increase Multiplicative Decrease}
\newacronym{am}{AM}{Acknowledged Mode}
\newacronym{amc}{AMC}{Adaptive Modulation and Coding}
\newacronym{amf}{AMF}{Access and Mobility Management Function}
\newacronym{aops}{AOPS}{Adaptive Order Prediction Scheduling}
\newacronym{api}{API}{Application Programming Interface}
\newacronym{apn}{APN}{Access Point Name}
\newacronym{aqm}{AQM}{Active Queue Management}
\newacronym{ausf}{AUSF}{Authentication Server Function}
\newacronym{avc}{AVC}{Advanced Video Coding}
\newacronym{awgn}{AGWN}{Additive White Gaussian Noise}
\newacronym{balia}{BALIA}{Balanced Link Adaptation Algorithm}
\newacronym{bbu}{BBU}{Base Band Unit}
\newacronym{bdp}{BDP}{Bandwidth-Delay Product}
\newacronym{ber}{BER}{Bit Error Rate}
\newacronym{bf}{BF}{Beamforming}
\newacronym{bler}{BLER}{Block Error Rate}
\newacronym{brr}{BRR}{Bayesian Ridge Regressor}
\newacronym{bsr}{BSR}{Buffer Status Report}
\newacronym{bs}{BS}{Base Station}
\newacronym{bss}{BSS}{Business Support System}
\newacronym{ca}{CA}{Carrier Aggregation}
\newacronym{caas}{CaaS}{Connectivity-as-a-Service}
\newacronym{cb}{CB}{Code Block}
\newacronym{cc}{CC}{Congestion Control}
\newacronym{ccid}{CCID}{Congestion Control ID}
\newacronym{cco}{CC}{Carrier Component}
\newacronym{cdd}{CDD}{Cyclic Delay Diversity}
\newacronym{cdf}{CDF}{Cumulative Distribution Function}
\newacronym{cdn}{CDN}{Content Distribution Network}
\newacronym{cir}{CIR}{Channel Impulse Response}
\newacronym{cn}{CN}{Core Network}
\newacronym{codel}{CoDel}{Controlled Delay Management}
\newacronym{comac}{COMAC}{Converged Multi-Access and Core}
\newacronym{cord}{CORD}{Central Office Re-architected as a Datacenter}
\newacronym{cornet}{CORNET}{COgnitive Radio NETwork}
\newacronym{cosmos}{COSMOS}{Cloud Enhanced Open Software Defined Mobile Wireless Testbed for City-Scale Deployment}
\newacronym{cots}{COTS}{Commercial Off-the-Shelf}
\newacronym{cp}{CP}{Control Plane}
\newacronym{cpu}{CPU}{Central Processing Unit}
\newacronym{cqi}{CQI}{Channel Quality Information}
\newacronym{cr}{CR}{Cognitive Radio}
\newacronym{cran}{CRAN}{Cloud \gls{ran}}
\newacronym{crs}{CRS}{Cell Reference Signal}
\newacronym{csi}{CSI}{Channel State Information}
\newacronym{csirs}{CSI-RS}{Channel State Information - Reference Signal}
\newacronym{cu}{CU}{Central Unit}
\newacronym{d2tcp}{D$^2$TCP}{Deadline-aware Data center TCP}
\newacronym{d3}{D$^3$}{Deadline-Driven Delivery}
\newacronym{dac}{DAC}{Digital to Analog Converter}
\newacronym{dag}{DAG}{Directed Acyclic Graph}
\newacronym{darpa}{DARPA}{Defense Advanced Research Projects Agency}
\newacronym{das}{DAS}{Distributed Antenna System}
\newacronym{dash}{DASH}{Dynamic Adaptive Streaming over HTTP}
\newacronym{dc}{DC}{Dual Connectivity}
\newacronym{dccp}{DCCP}{Datagram Congestion Control Protocol}
\newacronym{dce}{DCE}{Direct Code Execution}
\newacronym{dci}{DCI}{Downlink Control Information}
\newacronym{dcl}{DCL}{Dear Colleague Letter}
\newacronym{dctcp}{DCTCP}{Data Center TCP}
\newacronym{dl}{DL}{Downlink}
\newacronym{dmr}{DMR}{Deadline Miss Ratio}
\newacronym{dmrs}{DMRS}{DeModulation Reference Signal}
\newacronym{drlcc}{DRL-CC}{Deep Reinforcement Learning Congestion Control}
\newacronym{drs}{DRS}{Discovery Reference Signal}
\newacronym{du}{DU}{Distributed Unit}
\newacronym{e2e}{E2E}{end-to-end}
\newacronym{ecaas}{ECaaS}{Edge-Cloud-as-a-Service}
\newacronym{ecn}{ECN}{Explicit Congestion Notification}
\newacronym{edf}{EDF}{Earliest Deadline First}
\newacronym{embb}{eMBB}{Enhanced Mobile Broadband}
\newacronym{empower}{EMPOWER}{EMpowering transatlantic PlatfOrms for advanced WirEless Research}
\newacronym{enb}{eNB}{evolved Node Base}
\newacronym{endc}{EN-DC}{E-UTRAN-\gls{nr} \gls{dc}}
\newacronym{epc}{EPC}{Evolved Packet Core}
\newacronym{eps}{EPS}{Evolved Packet System}
\newacronym{es}{ES}{Edge Server}
\newacronym{etsi}{ETSI}{European Telecommunications Standards Institute}
\newacronym[firstplural=Estimated Times of Arrival (ETAs)]{eta}{ETA}{Estimated Time of Arrival}
\newacronym{eutran}{E-UTRAN}{Evolved Universal Terrestrial Access Network}
\newacronym{faas}{FaaS}{Function-as-a-Service}
\newacronym{fapi}{FAPI}{Functional Application Platform Interface}
\newacronym{fcc}{FCC}{Federal Communications Commission}
\newacronym{fdd}{FDD}{Frequency Division Duplexing}
\newacronym{fdm}{FDM}{Frequency Division Multiplexing}
\newacronym{fdma}{FDMA}{Frequency Division Multiple Access}
\newacronym{fed4fire}{FED4FIRE+}{Federation 4 Future Internet Research and Experimentation Plus}
\newacronym{fir}{FIR}{Finite Impulse Response}
\newacronym{fit}{FIT}{Future \acrlong{iot}}
\newacronym{fpga}{FPGA}{Field Programmable Gate Array}
\newacronym{fr2}{FR2}{Frequency Range 2}
\newacronym{fs}{FS}{Fast Switching}
\newacronym{fscc}{FSCC}{Flow Sharing Congestion Control}
\newacronym{ftp}{FTP}{File Transfer Protocol}
\newacronym{fw}{FW}{Flow Window}
\newacronym{ge}{GE}{Gaussian Elimination}
\newacronym{gnb}{gNB}{Next Generation Node Base}
\newacronym{gop}{GOP}{Group of Pictures}
\newacronym{gpr}{GPR}{Gaussian Process Regressor}
\newacronym{gpu}{GPU}{Graphics Processing Unit}
\newacronym{gtp}{GTP}{GPRS Tunneling Protocol}
\newacronym{gtpc}{GTP-C}{GPRS Tunnelling Protocol Control Plane}
\newacronym{gtpu}{GTP-U}{GPRS Tunnelling Protocol User Plane}
\newacronym{gtpv2c}{GTPv2-C}{\gls{gtp} v2 - Control}
\newacronym{gw}{GW}{Gateway}
\newacronym{harq}{HARQ}{Hybrid Automatic Repeat reQuest}
\newacronym{hetnet}{HetNet}{Heterogeneous Network}
\newacronym{hh}{HH}{Hard Handover}
\newacronym{hol}{HOL}{Head-of-Line}
\newacronym{hqf}{HQF}{Highest-quality-first}
\newacronym{hss}{HSS}{Home Subscription Server}
\newacronym{http}{HTTP}{HyperText Transfer Protocol}
\newacronym{ia}{IA}{Initial Access}
\newacronym{iab}{IAB}{Integrated Access and Backhaul}
\newacronym{ic}{IC}{Incident Command}
\newacronym{ietf}{IETF}{Internet Engineering Task Force}
\newacronym{imsi}{IMSI}{International Mobile Subscriber Identity}
\newacronym{imt}{IMT}{International Mobile Telecommunication}
\newacronym{iot}{IoT}{Internet of Things}
\newacronym{ip}{IP}{Internet Protocol}
\newacronym{itu}{ITU}{International Telecommunication Union}
\newacronym{kpi}{KPI}{Key Performance Indicator}
\newacronym{kvm}{KVM}{Kernel-based Virtual Machine}
\newacronym{los}{LOS}{Line-of-Sight}
\newacronym{lsm}{LSM}{Link-to-System Mapping}
\newacronym{lstm}{LSTM}{Long Short Term Memory}
\newacronym{lte}{LTE}{Long Term Evolution}
\newacronym{lxc}{LXC}{Linux Container}
\newacronym{m2m}{M2M}{Machine to Machine}
\newacronym{mac}{MAC}{Medium Access Control}
\newacronym{manet}{MANET}{Mobile Ad Hoc Network}
\newacronym{mano}{MANO}{Management and Orchestration}
\newacronym{mc}{MC}{Multi-Connectivity}
\newacronym{mcc}{MCC}{Mobile Cloud Computing}
\newacronym{mchem}{MCHEM}{Massive Channel Emulator}
\newacronym{mcs}{MCS}{Modulation and Coding Scheme}
\newacronym{mec}{MEC}{Multi-access Edge Computing}
\newacronym{mec2}{MEC}{Mobile Edge Cloud}
\newacronym{mfc}{MFC}{Mobile Fog Computing}
\newacronym{mi}{MI}{Mutual Information}
\newacronym{mib}{MIB}{Master Information Block}
\newacronym{miesm}{MIESM}{Mutual Information Based Effective SINR}
\newacronym{mimo}{MIMO}{Multiple Input, Multiple Output}
\newacronym{mgen}{MGEN}{Multi-Generator}
\newacronym{ml}{ML}{Machine Learning}
\newacronym{mlr}{MLR}{Maximum-local-rate}
\newacronym[plural=\gls{mme}s,firstplural=Mobility Management Entities (MMEs)]{mme}{MME}{Mobility Management Entity}
\newacronym{mmtc}{mMTC}{Massive Machine-Type Communications}
\newacronym{mmwave}{mmWave}{millimeter wave}
\newacronym{mpdccp}{MP-DCCP}{Multipath Datagram Congestion Control Protocol}
\newacronym{mptcp}{MPTCP}{Multipath TCP}
\newacronym{mr}{MR}{Maximum Rate}
\newacronym{mrdc}{MR-DC}{Multi \gls{rat} \gls{dc}}
\newacronym{mse}{MSE}{Mean Square Error}
\newacronym{mss}{MSS}{Maximum Segment Size}
\newacronym{mt}{MT}{Mobile Termination}
\newacronym{mtd}{MTD}{Machine-Type Device}
\newacronym{mtu}{MTU}{Maximum Transmission Unit}
\newacronym{mumimo}{MU-MIMO}{Multi-user \gls{mimo}}
\newacronym{mvno}{MVNO}{Mobile Virtual Network Operator}
\newacronym{nalu}{NALU}{Network Abstraction Layer Unit}
\newacronym{nas}{NAS}{Network Attached Storage}
\newacronym{nbiot}{NB-IoT}{Narrow Band IoT}
\newacronym{nfv}{NFV}{Network Function Virtualization}
\newacronym{nfvi}{NFVI}{Network Function Virtualization Infrastructure}
\newacronym{nic}{NIC}{Network Interface Card}
\newacronym{nlos}{NLOS}{Non-Line-of-Sight}
\newacronym{now}{NOW}{Non Overlapping Window}
\newacronym{nrdz}{NRDZ}{National Radio Dynamic Zone}
\newacronym{nsf}{NSF}{National Science Foundation}
\newacronym{nsm}{NSM}{Network Service Mesh}
\newacronym[type=hidden]{nr}{NR}{New Radio}
\newacronym{nrf}{NRF}{Network Repository Function}
\newacronym{nsa}{NSA}{Non Stand Alone}
\newacronym{nse}{NSE}{Network Slicing Engine}
\newacronym{nssf}{NSSF}{Network Slice Selection Function}
\newacronym{o2i}{O2I}{Outdoor to Indoor}
\newacronym{oai}{OAI}{OpenAirInterface}
\newacronym{oaicn}{OAI-CN}{\gls{oai} \acrlong{cn}}
\newacronym{oairan}{OAI-RAN}{\acrlong{oai} \acrlong{ran}}
\newacronym{oam}{OAM}{Operations, Administration and Maintenance}
\newacronym{ofdm}{OFDM}{Orthogonal Frequency Division Multiplexing}
\newacronym{olia}{OLIA}{Opportunistic Linked Increase Algorithm}
\newacronym{omec}{OMEC}{Open Mobile Evolved Core}
\newacronym{onap}{ONAP}{Open Network Automation Platform}
\newacronym{onf}{ONF}{Open Networking Foundation}
\newacronym{onos}{ONOS}{Open Networking Operating System}
\newacronym{oom}{OOM}{\gls{onap} Operations Manager}
\newacronym{opnfv}{OPNFV}{Open Platform for \gls{nfv}}
\newacronym[type=hidden]{oran}{O-RAN}{Open RAN}
\newacronym{orbit}{ORBIT}{Open-Access Research Testbed for Next-Generation Wireless Networks}
\newacronym{os}{OS}{Operating System}
\newacronym{oss}{OSS}{Operations Support System}
\newacronym{pa}{PA}{Position-aware}
\newacronym{pase}{PASE}{Prioritization, Arbitration, and Self-adjusting Endpoints}
\newacronym{pawr}{PAWR}{Platforms for Advanced Wireless Research}
\newacronym{pbch}{PBCH}{Physical Broadcast Channel}
\newacronym{pcef}{PCEF}{Policy and Charging Enforcement Function}
\newacronym{pcfich}{PCFICH}{Physical Control Format Indicator Channel}
\newacronym{pcrf}{PCRF}{Policy and Charging Rules Function}
\newacronym{pdcch}{PDCCH}{Physical Downlink Control Channel}
\newacronym{pdcp}{PDCP}{Packet Data Convergence Protocol}
\newacronym{pdsch}{PDSCH}{Physical Downlink Shared Channel}
\newacronym{pdu}{PDU}{Packet Data Unit}
\newacronym{pf}{PF}{Proportional Fair}
\newacronym{pgw}{PGW}{Packet Gateway}
\newacronym{phich}{PHICH}{Physical Hybrid ARQ Indicator Channel}
\newacronym{phy}{PHY}{Physical}
\newacronym{pmch}{PMCH}{Physical Multicast Channel}
\newacronym{pmi}{PMI}{Precoding Matrix Indicators}
\newacronym{powder}{POWDER}{Platform for Open Wireless Data-driven Experimental Research}
\newacronym{ppo}{PPO}{Proximal Policy Optimization}
\newacronym{ppp}{PPP}{Poisson Point Process}
\newacronym{prach}{PRACH}{Physical Random Access Channel}
\newacronym{prb}{PRB}{Physical Resource Block}
\newacronym{psnr}{PSNR}{Peak Signal to Noise Ratio}
\newacronym{pss}{PSS}{Primary Synchronization Signal}
\newacronym{pucch}{PUCCH}{Physical Uplink Control Channel}
\newacronym{pusch}{PUSCH}{Physical Uplink Shared Channel}
\newacronym{qam}{QAM}{Quadrature Amplitude Modulation}
\newacronym{qci}{QCI}{\gls{qos} Class Identifier}
\newacronym{qoe}{QoE}{Quality of Experience}
\newacronym{qos}{QoS}{Quality of Service}
\newacronym{quic}{QUIC}{Quick UDP Internet Connections}
\newacronym{rach}{RACH}{Random Access Channel}
\newacronym{ran}{RAN}{Radio Access Network}
\newacronym[firstplural=Radio Access Technologies (RATs)]{rat}{RAT}{Radio Access Technology}
\newacronym{rcn}{RCN}{Research Coordination Network}
\newacronym{rec}{REC}{Radio Edge Cloud}
\newacronym{red}{RED}{Random Early Detection}
\newacronym{renew}{RENEW}{Reconfigurable Eco-system for Next-generation End-to-end Wireless}
\newacronym{rf}{RF}{Radio Frequency}
\newacronym{rfc}{RFC}{Request for Comments}
\newacronym{rfr}{RFR}{Random Forest Regressor}
\newacronym{ric}{RIC}{RAN Intelligent Controller}
\newacronym{rlc}{RLC}{Radio Link Control}
\newacronym{rlf}{RLF}{Radio Link Failure}
\newacronym{rlnc}{RLNC}{Random Linear Network Coding}
\newacronym{rmse}{RMSE}{Root Mean Squared Error}
\newacronym{rnis}{RNIS}{Radio Network Information Service}
\newacronym{rr}{RR}{Round Robin}
\newacronym{rrc}{RRC}{Radio Resource Control}
\newacronym{rrm}{RRM}{Radio Resource Management}
\newacronym{rru}{RRU}{Remote Radio Unit}
\newacronym{rs}{RS}{Remote Server}
\newacronym{rsrp}{RSRP}{Reference Signal Received Power}
\newacronym{rsrq}{RSRQ}{Reference Signal Received Quality}
\newacronym{rss}{RSS}{Received Signal Strength}
\newacronym{rssi}{RSSI}{Received Signal Strength Indicator}
\newacronym{rtt}{RTT}{Round Trip Time}
\newacronym{ru}{RU}{Radio Unit}
\newacronym{rw}{RW}{Receive Window}
\newacronym{rx}{RX}{Receiver}
\newacronym{s1ap}{S1AP}{S1 Application Protocol}
\newacronym{sa}{SA}{standalone}
\newacronym{sack}{SACK}{Selective Acknowledgment}
\newacronym{sap}{SAP}{Service Access Point}
\newacronym{sc2}{SC2}{Spectrum Collaboration Challenge}
\newacronym{scef}{SCEF}{Service Capability Exposure Function}
\newacronym{sch}{SCH}{Secondary Cell Handover}
\newacronym{scoot}{SCOOT}{Split Cycle Offset Optimization Technique}
\newacronym{sctp}{SCTP}{Stream Control Transmission Protocol}
\newacronym{sdap}{SDAP}{Service Data Adaptation Protocol}
\newacronym{sdk}{SDK}{Software Development Kit}
\newacronym{sdm}{SDM}{Space Division Multiplexing}
\newacronym{sdma}{SDMA}{Spatial Division Multiple Access}
\newacronym{sdn}{SDN}{Software-defined Networking}
\newacronym{sdr}{SDR}{Software-defined Radio}
\newacronym{seba}{SEBA}{SDN-Enabled Broadband Access}
\newacronym{sgsn}{SGSN}{Serving GPRS Support Node}
\newacronym{sgw}{SGW}{Service Gateway}
\newacronym{si}{SI}{Study Item}
\newacronym{sib}{SIB}{Secondary Information Block}
\newacronym{sinr}{SINR}{Signal to Interference plus Noise Ratio}
\newacronym{sip}{SIP}{Session Initiation Protocol}
\newacronym{siso}{SISO}{Single Input, Single Output}
\newacronym{sla}{SLA}{Service Level Agreement}
\newacronym{sm}{SM}{Saturation Mode}
\newacronym{smf}{SMF}{Session Management Function}
\newacronym{smo}{SMO}{Service Management and Orchestration}
\newacronym{sms}{SMS}{Short Message Service}
\newacronym{smsgmsc}{SMS-GMSC}{\gls{sms}-Gateway}
\newacronym{snr}{SNR}{Signal-to-Noise-Ratio}
\newacronym{son}{SON}{Self-Organizing Network}
\newacronym{sptcp}{SPTCP}{Single Path TCP}
\newacronym{srb}{SRB}{Service Radio Bearer}
\newacronym{srn}{SRN}{Standard Radio Node}
\newacronym{srs}{SRS}{Sounding Reference Signal}
\newacronym{ss}{SS}{Synchronization Signal}
\newacronym{sss}{SSS}{Secondary Synchronization Signal}
\newacronym{st}{ST}{Spanning Tree}
\newacronym{svc}{SVC}{Scalable Video Coding}
\newacronym{tb}{TB}{Transport Block}
\newacronym{tdd}{TDD}{Time Division Duplexing}
\newacronym{tdm}{TDM}{Time Division Multiplexing}
\newacronym{tdma}{TDMA}{Time Division Multiple Access}
\newacronym{tfl}{TfL}{Transport for London}
\newacronym{tfrc}{TFRC}{TCP-Friendly Rate Control}
\newacronym{tft}{TFT}{Traffic Flow Template}
\newacronym{tgen}{TGEN}{Traffic Generator}
\newacronym{tip}{TIP}{Telecom Infra Project}
\newacronym{tm}{TM}{Transparent Mode}
\newacronym{to}{TO}{Telco Operator}
\newacronym{tr}{TR}{Technical Report}
\newacronym{trp}{TRP}{Transmitter Receiver Pair}
\newacronym{ts}{TS}{Technical Specification}
\newacronym{tti}{TTI}{Transmission Time Interval}
\newacronym{ttt}{TTT}{Time-to-Trigger}
\newacronym{tx}{TX}{Transmitter}
\newacronym{uas}{UAS}{Unmanned Aerial System}
\newacronym{uav}{UAV}{Unmanned Aerial Vehicle}
\newacronym{udm}{UDM}{Unified Data Management}

\newacronym{udr}{UDR}{Unified Data Repository}
\newacronym{ue}{UE}{User Equipment}
\newacronym{uhd}{UHD}{\gls{usrp} Hardware Driver}
\newacronym{ul}{UL}{Uplink}
\newacronym{um}{UM}{Unacknowledged Mode}
\newacronym{uml}{UML}{Unified Modeling Language}
\newacronym{upa}{UPA}{Uniform Planar Array}
\newacronym{upf}{UPF}{User Plane Function}
\newacronym{urllc}{URLLC}{Ultra Reliable and Low Latency Communication}
\newacronym{usa}{U.S.}{United States}
\newacronym{usim}{USIM}{Universal Subscriber Identity Module}
\newacronym{usrp}{USRP}{Universal Software Radio Peripheral}
\newacronym{utc}{UTC}{Urban Traffic Control}
\newacronym{vim}{VIM}{Virtualization Infrastructure Manager}
\newacronym{vm}{VM}{Virtual Machine}
\newacronym{vnf}{VNF}{Virtual Network Function}
\newacronym{volte}{VoLTE}{Voice over \gls{lte}}
\newacronym{voltha}{VOLTHA}{Virtual OLT HArdware Abstraction}
\newacronym{vr}{VR}{Virtual Reality}
\newacronym{vran}{vRAN}{Virtualized \gls{ran}}
\newacronym{vss}{VSS}{Video Streaming Server}
\newacronym{wbf}{WBF}{Wired Bias Function}
\newacronym{wf}{WF}{Wired-first}
\newacronym{wlan}{WLAN}{Wireless Local Area Network}
\newacronym{osm}{OSM}{Open Source \gls{nfv} Management and Orchestration}
\newacronym{pnf}{PNF}{Physical Network Function}
\newacronym{drl}{DRL}{Deep Reinforcement Learning}
\newacronym{mtc}{MTC}{Machine-type Communications}

\newacronym{ci}{CI}{cyberinfrastructure}
\newacronym{sonic}{SONiC}{Software for Open Networking in the Cloud}
\newacronym{ocp}{OCP}{Open Compute Project}
\newacronym{snmp}{SNMP}{Simple Network Management Protocol}
\newacronym{raid}{RAID}{redundant array of independent disks}
\newacronym{nfs}{NFS}{Network File Storage}
\newacronym{cin}{CI}{Continuous Integration}
\newacronym{cd}{CD}{Continuous Deployment}
\newacronym{dtn}{DTN}{Data Transfer Node}

\newacronym{dns}{DNS}{Domain Name Service}
\newacronym{nrpe}{NRPE}{Nagios Remote Plugin Executor}
\newacronym{ldap}{LDAP}{Lightweight Directory Access Protocol}
\newacronym{lan}{LAN}{Local Area Network}
\newacronym{vlan}{VLAN}{Virtual LAN}

\newacronym{ipmi}{IPMI}{Intelligent Platform Management Interface}
\newacronym{tor}{ToR}{Top-of-the-Rack}
\newacronym{lmn}{LMN}{Large Memory Node}
\newacronym{bgp}{BGP}{Border Gateway Protocol}
\newacronym{dhcp}{DHCP}{Dynamic Host Configuration Protocol}
\newacronym{vrf}{VRF}{Virtual Routing and Forwarding}
\newacronym{vpn}{VPN}{Virtual Private Network}
\newacronym{rma}{RMA}{Return Merchandise Authorization}
\newacronym{hpc}{HPC}{High Performance Compute}

\newacronym{nu}{NU}{Northeastern University}
\newacronym{asic}{ASIC}{Application-specific Integrated Circuit}
\newacronym{rdma}{RDMA}{Remote Direct Memory Access}
\newacronym{roce}{RoCE}{RDMA over Converged Ethernet}
\newacronym{ovs}{OVS}{Open vSwitch}
\newacronym{frr}{FRR}{Free Range Routing}
\newacronym{ups}{UPS}{Uninterruptible Power Supply}

\newacronym{ntia}{NTIA}{National Telecommunications and Information Administration}
\newacronym{pii}{PII}{Personal and Identifiable Information}
\newacronym{irb}{IRB}{Institutional Review Board}
\newacronym{doi}{DOI}{Digital Object Identifier}

\newacronym{sdo}{SDO}{Standard-Development Organization}
\newacronym{ose}{OSE}{Open Source Ecosystem}
\newacronym{osc}{OSC}{O-RAN Software Community}
\newacronym{dop}{DOP}{Director of Operations}
\newacronym{pm}{PM}{Program Manager}
\newacronym{excom}{EXCOM}{Executive Committee}
\newacronym{iiot}{IIoT}{Industrial \gls{iot}}
\newacronym{lf}{LF}{Linux Foundation}

\newacronym{6g}{6G}{Sixth-Generation}
\newacronym{a1}{A1}{O-RAN Policy Management Interface}
\newacronym{daa}{DAA}{Decentralized Authentication Authority}
\newacronym{dapp}{dApp}{Distributed RAN Application}
\newacronym{e2}{E2}{O-RAN Near-Real-Time Interface}
\newacronym{ecdf}{ECDF}{Empirical Cumulative Distribution Function}
\newacronym{fede2}{FED-E2}{Federated O-RAN Near-Real-Time Interface}
\newacronym{fl}{FL}{Feeder Link}
\newacronym{geo}{GEO}{Geostationary Orbit}
\newacronym{gnss}{GNSS}{Global Navigation Satellite System}
\newacronym{gs}{GS}{Geometric Spread}
\newacronym{hap}{HAP}{High Altitude Platform}
\newacronym{hdtn}{HDTN}{High-Delay Tolerant Networking}
\newacronym{iq}{IQ}{In-phase and Quadrature}
\newacronym{isl}{ISL}{Inter-Satellite Link}
\newacronym{leo}{LEO}{Low-Earth Orbit}
\newacronym{meo}{MEO}{Medium Earth Orbit}
\newacronym{mppt}{MPPT}{ Maximum Power Point Tracking }
\newacronym{ntn}{NTN}{Non-Terrestrial Network}
\newacronym{o1}{O1}{O-RAN Operations and Management Interface}
\newacronym{osiran}{OSIRAN}{Open-Space-Integrated-RAN}
\newacronym{pe}{PE}{Pointing Error}
\newacronym{pdf}{PDF}{Probability Density Function}
\newacronym{rapp}{rApp}{RAN Application}
\newacronym{sapp}{sApp}{Space Application}
\newacronym{sagin}{SAGIN}{Space-Air-Ground Integrated Network}
\newacronym{sl}{SL}{Service Link}
\newacronym{spaceran}{Space-RAN}{Space integrated Open Radio Access network}
\newacronym{spaceric}{Space-RIC}{Space RAN Intelligent Controller}
\newacronym{sru}{s-RU}{Satellite Radio Unit}
\newacronym{sdu}{s-DU}{Satellite Distributed Unit}
\newacronym{scu}{s-CU}{Satellite Central Unit}
\newacronym{snf}{s-NF}{Satellite Network Function}
\newacronym{tn}{TN}{Terrestrial Network}
\newacronym{tle}{TLE}{Two-Line Element Set}
\newacronym{gsl}{GSL}{ground-to-satellite link}

\newacronym{fso}{FSO}{Free-Space Optics}
\newacronym{circ}{CIRC}{Community Infrastructure for Research in Computer and Information Science and Engineering}
\newacronym{trl}{TRL}{Technology Readiness Level}

\newacronym{isac}{ISAC}{Integrated Sensing and Communications}
\newacronym{dass}{DASS}{Dynamically Adjustable Spectrum Sharing between Ground Communication Networks and Earth Exploration Satellite Systems Above 100~GHz}
\newacronym{stk}{STK}{System Tool Kit}
\newacronym{ris}{RIS}{Reconfigurable Intelligent Surface}
\newacronym{noma}{NOMA}{Non-Orthogonal Multiple Access}

\newacronym{vipr}{VIPR}{Vapor In-cloud Profiling Radar}
\newacronym{dar}{DAR}{Differential Absorption Radar}
\newacronym{dial}{DIAL}{Differential Absorption LIDAR}
\newacronym{scd}{SCD}{Sensing Centric Design}
\newacronym{ccd}{CCD}{Communications Centric Design}
\newacronym{jd}{JD}{Joint Design}
\newacronym{sar}{SAR}{Synthetic Aperture Radar}
\newacronym{fmcw}{FMCW}{Frequency Modulated Continuous Wave}
\newacronym{gml}{GML}{Generalized Maximum Likelihood}
\newacronym{hitran}{HITRAN}{HIgh resolution TRANsmission molecular absorption}
\newacronym{mae}{MAE}{Mean Absolute Error}
\newacronym{psd}{PSD}{Power Spectral Density}
\newacronym{css}{CSS}{chirp spread spectrum}
\newacronym{eess}{EESS}{Earth exploration satellite services}

\newacronym{CFO}{CFO}{carrier frequency offset}
\newacronym{CR}{CR}{clock recovery}
\newacronym{BB}{BB}{baseband}
\newacronym{QAM}{QAM}{quadrature amplitude modulation}
\newacronym{IF}{IF}{intermediate frequency}
\newacronym{IQ}{IQ}{in-phase and quadrature}
\newacronym{LO}{LO}{local oscillator}
\newacronym{PN}{PN}{phase noise}
\newacronym{TED}{TED}{timing error detector}
\newacronym{CMA}{CMA}{constant modulus algorithm}
\newacronym{DPLL}{DPLL}{digital phase-locked loop}
\newacronym{DD}{DD}{decision directed}
\newacronym{NCO}{NCO}{numerically controlled oscillator}
\newacronym{ISI}{ISI}{intersymbol interference}
\newacronym{RF}{RF}{radio frequency}
\newacronym{BER}{BER}{bit error rate}
\newacronym{EVM}{EVM}{error vector magnitude}
\newacronym{DMIMO}{DMIMO}{distributed multiple-input multiple-output}
\newacronym{MIMO}{MIMO}{multiple input multiple output}
\newacronym{PSG}{PSG}{performance signal generator}
\newacronym{AWG}{AWG}{arbitrary waveform generator}
\newacronym{DSO}{DSO}{digital storage oscilloscope}
\newacronym{PA}{PA}{power amplifier}
\newacronym{LNA}{LNA}{low-noise amplifier}
\newacronym{AMC}{AMC}{active multiplier chain}
\newacronym{DSP}{DSP}{digital signal processing}
\newacronym{ML}{ML}{maximum likelihood}
\newacronym{AWGN}{AWGN}{additive white Gaussian noise}
\newacronym{PLL}{PLL}{phase-locked loop}
\newacronym{OEPLL}{OEPLL}{optoelectronic phase-locked loop}
\newacronym{sthz}{sub-THz}{sub-terahertz}
\newacronym{thz}{THz}{terahertz}
\newacronym{RRC}{RRC}{root raised cosine}
\newacronym{papr}{PAPR}{Peak-to-Average Power Ratio}
\newacronym{iss}{ISS}{International Space Station}
\newacronym{dpu}{DPU}{Data Processing Unit}
\newacronym{ce}{CE}{Channel Equalization}
\newacronym{fft}{FFT}{Fast Fourier Transform}
\newacronym{genai}{GenAI}{Generative Artificial Intelligence}
\newacronym{ddos}{DDoS}{Distributed Denial of Service}

\newacronym{tcp}{TCP}{Transmission Control Protocol}
\newacronym{udp}{UDP}{User Datagram Protocol}
\newacronym{tls}{TLS}{Transport Layer Security}
\newacronym{ssl}{SSL}{Secure Sockets Layer}
\newacronym{mdns}{mDNS}{Multicast Domain Name System}
\newacronym{mqtt}{MQTT}{Message Queuing Telemetry Transport}
\newacronym{stun}{STUN}{Session Traversal Utilities for NAT}
\newacronym{ntp}{NTP}{Network Time Protocol}
\newacronym{bootp}{BOOTP}{Bootstrap Protocol}
\newacronym{upnp}{UPnP}{Universal Plug and Play}
\newacronym{ssdp}{SSDP}{Simple Service Discovery Protocol}
\newacronym{coap}{CoAP}{Constrained Application Protocol}
\newacronym{nat}{NAT}{Network Address Translation}
\newacronym{ap}{AP}{Access Point}
\newacronym{wan}{WAN}{Wide Area Network}
\newacronym{ota}{OTA}{Over-the-Air}
\newacronym{wpa}{WPA}{Wi-Fi Protected Access}
\newacronym{isp}{ISP}{Internet Service Provider}
\newacronym{rest}{REST}{Representational State Transfer}
\newacronym{p4}{P4}{Programming Protocol-independent Packet Processors}
\newacronym{dfa}{DFA}{Deterministic Finite Automaton}
\newacronym{fsm}{FSM}{Finite State Machine}
\newacronym{e2ap}{E2AP}{E2 Application Protocol}
\newacronym{ble}{BLE}{Bluetooth Low Energy}
\newacronym{zigbee}{Zigbee}{Zigbee Protocol}
\newacronym{zwave}{Z-Wave}{Z-Wave Protocol}
\newacronym{svm}{SVM}{Support Vector Machine}
\newacronym{knn}{k-NN}{k-Nearest Neighbors}
\newacronym{dt}{DT}{Decision Tree}
\newacronym{mlp}{MLP}{Multi-Layer Perceptron}
\newacronym{gru}{GRU}{Gated Recurrent Unit}
\newacronym{bilstm}{BiLSTM}{Bidirectional Long Short-Term Memory}
\newacronym{gb}{GB}{Gradient Boosting}
\newacronym{gbc}{GBC}{Gradient Boosting Classifier}
\newacronym{svd}{SVD}{Singular Value Decomposition}
\newacronym{tsne}{t-SNE}{t-distributed Stochastic Neighbor Embedding}
\newacronym{dbscan}{DBSCAN}{Density-Based Spatial Clustering of Applications with Noise}
\newacronym{nlp}{NLP}{Natural Language Processing}

\newacronym{wifi}{Wi-Fi}{Wireless Fidelity (IEEE 802.11)}
\newacronym{lora}{LoRa}{Long Range}
\newacronym{ram}{RAM}{Random Access Memory}
\newacronym{llm}{LLM}{Large Language Model}
\newacronym{nn}{NN}{Neural Network}
\newacronym{pca}{PCA}{Principal Component Analysis}
\newacronym{cnn}{CNN}{Convolutional Neural Network}
\newacronym{rnn}{RNN}{Recurrent Neural Network}
\newacronym{hmm}{HMM}{Hidden Markov Model}
\newacronym{fhmm}{FHMM}{Factorial Hidden Markov Model}
\newacronym{dft}{DFT}{Discrete Fourier Transform}
\newacronym{dpi}{DPI}{Deep Packet Inspection}
\newacronym{tee}{TEE}{Trusted Execution Environment}
\newacronym{acl}{ACL}{Access Control List}

\usepackage[many]{tcolorbox}

 \newcommand{\findingsbox}[1]{
\begin{tcolorbox}[breakable,width=\linewidth,
boxrule=0pt, leftrule = 6pt, top=1pt, bottom=1pt, left=1pt,right=1pt, 
colback=gray!20,colframe=gray!60]
\textbf{Takeaway Message:} #1
\end{tcolorbox}
}

\begin{document}
\title{ A Comprehensive Survey on Smart Home IoT Fingerprinting: From Detection to Prevention and Practical Deployment }

\author{
\IEEEauthorblockN{
Eduardo Baena\IEEEauthorrefmark{2},
Han Yang\IEEEauthorrefmark{1}, 
Dimitrios Koutsonikolas\IEEEauthorrefmark{2},
and
Israat Haque\IEEEauthorrefmark{1}\\
}  

\IEEEauthorblockA{\IEEEauthorrefmark{1}Dalhousie University, Canada; \IEEEauthorrefmark{2}Northeastern University, USA }
}

%\title{Mapping IoT Fingerprinting Literature: A Systematic Review}
%\author{, , , Israat Haque}
%\affil{ Faculty of Computer Science, Dalhousie University}

\maketitle

\begin{abstract}

Smart homes are increasingly populated with heterogeneous \gls{iot} devices that interact continuously with users and the environment. This diversity introduces critical challenges in device identification, authentication, and security, where fingerprinting techniques have emerged as a key approach. In this survey, we provide a comprehensive analysis of IoT fingerprinting specifically in the context of smart homes, examining methods for device and their event detection, classification, and intrusion prevention. We review existing techniques, e.g., network traffic analysis or machine learning–based schemes, highlighting their applicability and limitations in home environments characterized by resource-constrained devices, dynamic usage patterns, and privacy requirements. Furthermore, we discuss fingerprinting system deployment challenges like scalability, interoperability, and energy efficiency, as well as emerging opportunities enabled by generative \gls{ai} and federated learning. Finally, we outline open research directions that can advance reliable and privacy-preserving fingerprinting for next-generation smart home ecosystems.

%The number of \gls{iot} devices connected to the internet has grown exponentially over the last years. Unfortunately, the security and privacy concerns surrounding them has grown at a similar rate. To protect these devices and their users, the first natural step for a network operator is to fingerprint them, i.e., obtain information about the device manufacturer, type, expected behavior and/or user activity. At the same time, it is important to prevent malicious actors from being able to access the same information. In this paper, we provide the first comprehensive survey of the state-of-the-art in the \gls{iot} device fingerprinting domain. We start by covering the main fingerprinting techniques, how they work, their major design aspects and intended purpose. Next, we systematize existing defenses to prevent \gls{iot} device fingerprinting. Lastly, we detail current research challenges and identify areas for future development. While providing an extensive analysis, this survey can also be used as an accessible entry point to newcomers in the field.

\end{abstract}

\begin{IEEEkeywords}
Internet of Things, fingerprinting, network traffic analysis, security, privacy, smart home.
\end{IEEEkeywords}

%\begin{picture}(0,0)(0,-430)
%  \put(20,0){
%    \setlength{\fboxsep}{5pt} % Espacio interno del recuadro
%    \fbox{
%      \begin{minipage}{0.8\textwidth}
%        \footnotesize
%        \centering
%This work has been submitted to the IEEE for possible publication. \\
%Copyright may be transferred without notice, after which this version may no longer be %accessible.
%      \end{minipage}
%    }
%  }
%\end{picture}

\section{Introduction} \label{Intro}

%The rapid expansion of the \gls{iot} is transforming everyday life, with projections indicating that the number of connected devices will exceed 65 billion by 2025 \cite{smartcitiesandcampuses}. While \gls{iot} technologies are being deployed across diverse domains such as smart cities, industrial automation, and intelligent transportation, the latter remains the most prominent, accounting for nearly half of all \gls{iot} deployments \cite{cisco-report}. 

The rapid expansion of the \gls{iot} is transforming everyday life, with recent estimates projecting that the number of connected devices will surpass 40 billion by 2030 \cite{dhull2024iot}. Among the diverse application domains of \gls{iot}, smart homes have emerged as one of the fastest growing and most pervasive environments, where connected devices directly integrate into daily human activities and private spaces \cite{verizon2025connections}. This widespread adoption is due to the convenience and automation that \gls{iot} applications offer in everyday life. However, such massive growth and adoption expose users and service providers to security and privacy risks stemming from the presence of heterogeneous devices. Smart home \gls{iot} devices range from smart cameras and voice assistants to thermostats and lighting systems manufactured by vendors like Amazon, Google, Samsung, and D-Link, each with varying security protocols. Hence, intruders can exploit a growing number of vulnerabilities due to the protocol inconsistency and the level of defence mechanisms. For instance, the number of \gls{iot} devices involved in botnet-driven \gls{ddos} attacks has soared from approximately 200,000 to nearly 1 million in just one year \cite{nokia2023}. Such statistics highlight the critical need for robust security measures to protect these devices and the privacy of their users.

\begin{figure} [h]
\centering
\includegraphics[width=2.5in]{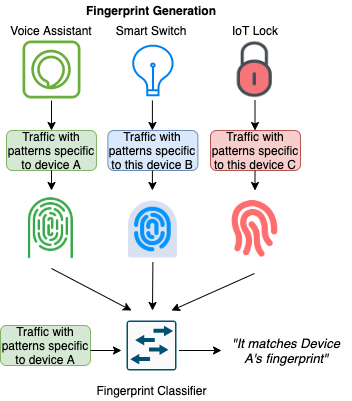}
\caption{Each fingerprint captures the essence of network traffic characteristics specific to its respective device.}
\label{fig:idea}
\end{figure}

A typical smart home \gls{iot} platform consists of: (a) \gls{iot} devices, (b) control applications that enable users to manage and interact with devices, and (c) cloud-based back-end services responsible for data storage, processing, and synchronization across locations. 
These \gls{iot} devices are typically installed and maintained by end-users, who may lack the technical expertise to implement advanced security measures unlike industrial or municipal \gls{iot} applications, where devices are often managed by skilled professionals. Thus, attackers particularly find smart homes as sweet targets to acquire sensitive information.
One effective approach to securing \gls{iot} devices involves monitoring their communication patterns for anomalies and generating alerts when suspicious behavior is detected. This method of identifying a device or its activity based on traffic characteristics is known as \textit{fingerprinting}. As shown in Fig.~\ref{fig:idea}, the fingerprinting process captures distinct traffic patterns from devices like voice assistants, smart switches, and \gls{iot} locks, which are then classified to identify the corresponding device. This capability allows for the detection of anomalies, enforcement of security policies, and prevention of unauthorized access. The sporadic nature of \gls{iot} communication makes network traffic-based fingerprinting a popular, accurate, and reliable method at scale. However, this traffic is usually encrypted; thus, learning-based passive analysis, which examines packet headers and metadata, is commonly used for encrypted traffic analysis \cite{liu, 2003detection, 1991prevention}.
%However, it also presents a potential risk if exploited by malicious actors to gain insights into device behavior or infer sensitive information about users. Although such leakage has a negative impact on the security and privacy of any user, they suffer the most in smart homes due to direct usage and interactions with devices.   

Numerous studies have explored the realm of smart home \gls{iot} traffic fingerprinting, which can be broadly classified into two primary categories: \textit{fingerprint detection} and \textit{fingerprint prevention}. Fingerprint detection focuses on identifying devices and their operational states, e.g., device identification and abnormal behavior detection, which can be invaluable for service providers and manufacturers in pinpointing device malfunctions or misbehavior, allowing them to take timely corrective actions \cite{automaticclass}. For home users, these techniques offer the ability to monitor and report any unusual device activity to relevant service providers, enhancing the security of their smart environments \cite{homesnitch}. Moreover, fingerprint detection extends beyond security monitoring; it can be employed in other contexts, such as distinguishing between different voice commands issued to smart assistants, thereby improving the functionality and user experience of such devices \cite{voicetraffic}. However, the same capabilities that enable these benefits also present risks, as adversaries could potentially exploit fingerprinting techniques to hijack devices \cite{zhouDiscover2019, chenIoTMine2019} or infer sensitive details about users' daily routines \cite{zhouIoTbeholder2023}. This is where fingerprint prevention mechanisms become crucial, as they aim to protect \gls{iot} devices and their traffic from such malicious exploitation. Techniques such as traffic padding are designed to obscure the traffic patterns that could otherwise be used to compromise device security.

\textbf{Contribution:} Given the critical role of both detection and prevention in maintaining the security of smart home \gls{iot} environments, a comprehensive review of these mechanisms is essential. Such a review would not only enhance our understanding of current approaches but also guide the development of more robust solutions. However, the existing literature often addresses these aspects in isolation, lacking a unified perspective. Previous surveys address either detection \cite{safi_survey_2022, chowdhury_survey_2022} or prevention without their deployment feasibility \cite{jmila_survey_2022}.  We assess the feasibility of deploying these techniques in real-world scenarios, considering factors such as data collection, feature selection, and the practical implementation of fingerprinting methods. Furthermore, we explore the transformative potential of \gls{genai} in improving \gls{iot} fingerprinting, identifying existing shortcomings, and suggesting future research directions. Despite extensive theoretical research on \gls{iot} fingerprinting, a gap remains regarding its deployment feasibility. This survey systematically examines this disconnect and outlines paths toward practical realization. 

\textbf{Organization:} Section~\ref{sec:scope} outlines the scope of the survey, focusing on smart home \gls{iot} devices and their network traffic-based fingerprinting. Section~\ref{sec:methodology} presents our methodology for paper collection and filtering. Section~\ref{sec:related work} discusses related surveys and delineates the novelty of our work. Section~\ref{sec: detection} examines techniques for extracting device-related information, including device identification~\ref{sec: discovering}, activity inference~\ref{sec: infer}, and policy enforcement~\ref{sec: policies}. Section~\ref{sec:prevention} explores methods for preventing fingerprinting, covering packet padding~\ref{sec: padding}, traffic injection~\ref{sec: injection}, traffic shaping~\ref{sec: shaping}, and other/hybrid obfuscation techniques~\ref{sec:hybrid}. Both sections address machine learning (ML) and non-ML methods with attention to deployment feasibility in smart homes. Section~\ref{sec:genai} provides a dedicated analysis of GenAI-based detection and prevention, emphasizing their practical potential. Section~\ref{sec: challenge} outlines key research challenges and future directions. Finally, Section~\ref{sec: conclusion} concludes with the main findings and implications of this survey.

\section{Scope and Methodology}

This section first positions the scope of the article and then presents the paper selection criteria.

\subsection{Scope of the Survey}
\label{sec:scope}

%\subsubsection{IoT application domain}

This study investigates smart home IoT devices,including personal wearables and home-based systems,as they directly interface with end-users and pose significant implications for privacy and network security.  We specifically target the most commonly deployed network traffic-based fingerprinting leveraging standard communication protocols, such as \gls{wifi}, Bluetooth, \gls{ble}, ZigBee, and \gls{lora}.  We explore both detection and prevention mechanisms, where the former schemes are categorized into \textit{device identification}, \textit{abnormal behavior detection}, \textit{device activity inference}, and \textit{policy enforcement}. Prevention schemes are classified into \textit{packet padding}, \textit{traffic injection}, and \textit{traffic shaping}. The novelty of this study lies in its comprehensive assessment of the deployment feasibility of fingerprinting schemes in smart homes, focusing on two critical dimensions: (1) the detection and prevention methods employed, and (2) the selection and utilization of data and features in data-driven solutions.

We consider both the \textit{accessibility} and \textit{applicability} of data collection platforms. Accessibility refers to the practical availability of platforms capable of capturing the required data,such as network adapters compatible with IEEE 802.11 standards. Applicability pertains to the quality and relevance of the collected data, considering factors such as device diversity, duration of data collection, and the context in which the data is gathered (e.g., simulated environments, controlled lab settings, or real-world deployments). Furthermore, our investigation includes both encrypted and unencrypted communications from different network layers, recognizing that the complexity of fingerprinting can vary based on specific protocols and encryption methods.  

%We distinguish these network traffic between packets and flows, identified using 5-tuple (source IP, destination IP, source port, destination port, protocol), traverse between IoT devices and their cloud servers. 

We categorize network traffic into packets and flows. Flows are identified using the 5-tuple (source \gls{ip} address, destination \gls{ip} address, source port, destination port, and protocol), and they represent the communication between \gls{iot} devices and their cloud servers. The analyzed traffic may carry \textit{time domain} features like packet inter-arrival times and the number of packets per unit of time useful for pattern recognition and anomaly detections \cite{meidan2018nbaiot, shafiq2020iot}. It is also possible to conduct \textit{frequency domain} analysis of the traffic using \gls{fft}~\cite{fftCirillo} or wavelet \cite{wavelet} to distinguish between control and event traffic. The collected features are the foundation of developing learning-based or data-driven fingerprinting with automation, adaptability, and scalability advantages over non-learning-based methods \cite{8887356}. 

However, these benefits of learning-based schemes often come with high resource demands, such as increased computation and memory requirements. Additionally, supervised learning schemes incur extra costs due to the need for data labeling. In contrast, non-learning-based fingerprinting methods provide several advantages, including lower resource requirements, simplicity, interpretability, the absence of a need for extensive training data, deterministic outcomes, and faster implementation \cite{Baral23}. For learning-based methods, we examine their data collection and labeling requirements, feature selection, and model selection criteria concerning deployment feasibility. For non-learning-based approaches, we review their fingerprinting algorithms and assess the feasibility of their deployment, focusing on the types of network data they utilize and their implementation requirements in real-world settings.

In summary, this survey centers on fingerprinting techniques that examine both encrypted and non-encrypted network traffic, with a focus on approaches that are feasible to deploy in real-world smart home settings.

\subsection{Methodology} \label{sec:methodology}

This section presents the methodology of selecting existing literature following the scope of the article. We define inclusion criteria (IC), exclusion criteria (EC), and searching keywords to explore academic repositories and select articles that meet the selection criteria.

\subsubsection{Paper Selection}

This section elaborates on the process of selecting papers. We define a search string based on keywords and synonyms related to the research scope to find relevant papers. The keywords (Table \ref{table:met_search}) include four groups: 
%\textcolor{red}{three groups}: 1) smart home IoT domain that includes appliances and personal IoT devices, 2) device-related terms, and 3) schemes that deploy fingerprint detection or prevention techniques. 

1) smart home IoT domain that includes appliances and personal IoT devices, 
2) communication and behavioral features (e.g., traffic, protocol, behavior), 
3) schemes that deploy fingerprinting detection or prevention techniques (e.g., monitor, profiling, padding), and 
4) the specific target entity (e.g., device instance, user activity, side channel). 

The search requires that at least one term from each of the four groups appear in the abstract for a paper to be included. This ensures the paper describes a relevant domain, includes a feature or communication characteristic, and involves a detection or protection mechanism applied to a specific target. The Detection/Protection column in the table lists terms related to either detection (marked in blue) or protection/prevention (marked in purple). If a term is associated with detection, such as fingerprint, profiling, or monitoring, we searched for these terms. Similarly, for protection or prevention, such as obfuscation, padding, or injection, we included these terms in the search. We also considered singular and plural forms, verb and noun forms, and other natural variations of these terms to ensure comprehensive coverage (e.g., flow and flows, detect and detection, monitor and monitoring). We searched the IEEE and ACM digital libraries using the constructed search terms and rules described above. Table~\ref{table:search_str} shows an example of an exact search string applied.
%\textcolor{red}{which is correct? --- four groups} CLARYFIED

\begin{table}[h]
    \centering
    \caption{Terms and synonyms used during the search.}
    \label{table:met_search}
    \begin{tabular}{|c|c|c|c|}
        \hline
        \textbf{Domain} & \textbf{Feature} & \textcolor{blue}{\textbf{Detection}}/\textcolor{purple}{\textbf{Protection}}& \textbf{Target} \\
        \hline
        IoT & traffic & \textcolor{blue}{fingerprint}/\textcolor{purple}{obfuscation} & device instance \\
        \hline
        smart home & flow & \textcolor{blue}{profiling}/\textcolor{purple}{padding} & device model \\
        \hline
        & behavior & \textcolor{blue}{identify}/\textcolor{purple}{shaping} & device type \\
        \hline
        & network & \textcolor{blue}{detect}/\textcolor{purple}{prevent} & device policy \\
        \hline
        & protocol & \textcolor{blue}{monitor}/\textcolor{purple}{injection} & device event \\
        \hline
        & & \textcolor{blue}{infer} & user activity \\
        \hline
        & & \textcolor{blue}{classify} & device state\\
        \hline
        & & & side channel \\
        \hline
    \end{tabular}
\end{table}

\begin{table}[h]
    \centering
    \caption{Search strings}
    \label{table:search_str}
    \begin{tabularx}{\linewidth}{X}
        \hline
((("Abstract":"IoT") OR ("Abstract":"smart home")) AND (("Abstract":"traffic*") OR ("Abstract":"packet*") OR ("Abstract":"flow*") OR ("Abstract":"behavior*") OR ("Abstract":"network") OR ("Abstract":"protocol")) AND (("Abstract":"fingerprint*") OR ("Abstract":"profil*") OR ("Abstract":"identi*") OR ("Abstract":"detect*") OR ("Abstract":"monit*") OR ("Abstract":"infer*") OR ("Abstract":"classify")) AND (("Abstract":"device instance") OR ("Abstract":"device model") OR ("Abstract":"device type") OR ("Abstract":"device policy") OR ("Abstract":"device event") OR ("Abstract":"user activi*") OR ("Abstract":"device state*") OR ("Abstract":"side channel")))\\
        \hline
    \end{tabularx}
\end{table}

To further refine the retrieved papers to fit our defined scope, Table \ref{table:inc_exc} outlines the inclusion criteria (IC) and exclusion criteria (EC) used to filter the most relevant studies from the initial pool of search results. Our selection is guided by three key focus areas: the smart home IoT domain, network traffic as the primary feature set, and detection/prevention as the main objectives. Given the rapid evolution of smart home devices, and the potential obsolescence of older techniques, we limit our review to studies published within the last 10 years (post-2014). Each paper is first evaluated against the inclusion criteria and then against the exclusion criteria. A study is retained only if it meets all ICs and does not fall under any ECs.

Upon a thorough analysis of these filtered papers, along with their relevant citations and references, we observed that some pertinent papers were not captured by the initial search strings but are relevant to the scope of this article. To address this, we manually reviewed the retrieved papers, cross-checked their citations, and added additional relevant works to our pool, including papers from arXiv and other sources.

\begin{table}[h]
    \centering
    \caption{Inclusion and Exclusion Criteria. }
    \label{table:inc_exc}
    \begin{tabular}{|c|p{0.7\linewidth}|}
        %\textbf{Code} & \textbf{Description} \\
        \hline
        \multicolumn{2}{|c|}{Inclusion Criteria} \\ 
        \hline
        IC 1 & The paper utilizes network traffic for analysis. \\
        \hline
        IC 2 & The paper implements fingerprinting in the IoT application domain. \\
        \hline
        IC 3 & The paper covers device detection such as identification, anomaly detection, device activity inference, and policy enforcement. \\
        \hline
        IC 4 & The paper covers fingerprinting obfuscation.\\
        \hline

        \hline
        \multicolumn{2}{|c|}{Exclusion Criteria} \\
        \hline
        EC 1 & The paper uses features at the physical layer (e.g., RF signal).\\
        \hline
        EC 2 & The paper takes non-fingerprinting approaches to perform device identification (e.g., MAC address parsing, prob request sends).\\
        \hline
        EC 3 & Fingerprinting is not the main goal of the paper. \\
        \hline
        EC 4 & The paper was published before 2014. \\
        \hline
    \end{tabular}
\end{table}

%\textcolor{blue}{As a result, after filtering, we identified a total of X\_detect = 501 detection papers from our initial search. Based on cross-citation checks, we manually added Y\_detect = 30 additional detection papers, yielding a final total of Z\_detect = 531 detection papers. Similarly, for prevention papers, we obtained X\_prevention = 23 papers through filtering the initial search and added Y\_prevention = 15 papers through cross-citation, resulting in a final pool of Z\_prevention = 38 prevention papers.}

After filtering the initial search results, we identified 501 detection papers. An additional 30 papers were included through cross-citation checks, leading to a final set of 531 detection papers. For prevention-related works, we found 23 papers through the initial search and identified 15 more via cross-citation, for a total of 38 prevention papers.

\section{Related Work} \label{sec:related work}

Table~\ref{table:related_work} presents existing surveys on \gls{iot} fingerprinting that we compare against our objectives of fingerprinting detection, prevention, and deployment feasibility to position the novelty of our contribution. Then, we outline two identified research gaps that we fill as the key contribution of this article.
The table shows that while there are several surveys on fingerprinting detection schemes, such as device identification and event inference  \cite{sanchez, safi_survey_2022, chowdhury_survey_2022, jmila_survey_2022}, these studies often overlook important aspects. For instance, a few surveys touch on IoT anomaly detection and policy enforcement \cite{chowdhury_survey_2022, safi_survey_2022}, but they do not specifically focus on smart home \gls{iot} devices. Moreover, none of these surveys thoroughly addresses prevention schemes, which are critical for both users and service providers.

\begin{table*}[t]
\centering
\fontsize{8pt}{10pt}\selectfont
\caption{Comparison with existing surveys.}
\label{table:related_work}
\begin{tabular}{@{}ccccccc@{}}
\toprule
\multirow{2}{*}{\textbf{Work}} & \multirow{2}{*}{\textbf{Year}} & \multicolumn{1}{c}{Domain} & \multicolumn{2}{c}{Detection-Prevention Approach} & \multicolumn{1}{c}{Deployment} & \multicolumn{1}{c}{Gen}\\

\cmidrule(lr){3-3} \cmidrule(lr){4-5} \cmidrule(lr){6-6} \cmidrule(lr){7-7}
& & \centering smart home Oriented & \centering Detection & \centering Prevention & Feasibility & AI\\ 
\midrule

Baldini et al. \cite{baldini-survey} & 2017 & \ding{55} & \(\bigcirc\) & \ding{55} & \ding{55} & \ding{55}\\

Yadav et al. \cite{yadav} & 2020 & \ding{55} & \(\bigcirc\) & \ding{55} & \ding{55} & \ding{55}\\

Sánchez et al. \cite{sanchez} & 2021 & \ding{55} & \(\bigcirc\) & \ding{55} & \ding{55} & \ding{55}\\

Miraqa Safi et al. \cite{safi_survey_2022} & 2022 & \ding{55} & \checkmark & \ding{55} & \ding{55} & \ding{55}\\

R.R. Chowdhury et al. \cite{chowdhury_survey_2022} & 2022 & \ding{55} & \checkmark & \ding{55} & \ding{55} & \ding{55}\\

H. Jmila et al. \cite{jmila_survey_2022} & 2022 & \checkmark & \(\bigcirc\) & \ding{55} & \ding{55} & \ding{55}\\

Our work & 2025 & \checkmark & \checkmark & \checkmark & \checkmark & \checkmark\\

\bottomrule
\end{tabular}
\medskip
\footnotesize
\begin{tabular}{@{}l@{}}
\checkmark: Full Coverage; \ding{55}: Not Included; \(\bigcirc\): Partial Coverage
\end{tabular}

\end{table*}

Another major gap in the existing literature is the lack of attention to the practical feasibility of deploying fingerprinting systems. This is a crucial consideration for users who need reliable methods to protect their smart homes. While some surveys \cite{safi_survey_2022, chowdhury_survey_2022, jmila_survey_2022} discuss fingerprinting datasets and commonly used features, they fail to evaluate whether these approaches can be effectively deployed in real-world scenarios. Our survey addresses this gap by examining the deployment aspects, including the platforms, tools, and techniques necessary for implementing fingerprinting systems. We also consider how these systems can handle different types of data, features, and algorithms. Furthermore, with the advent of \gls{genai}, new opportunities are emerging in the field of IoT fingerprinting, such as the creation of synthetic data and the automation of fingerprinting processes. Existing surveys have not yet explored this promising technology, which our article integrates to set the stage for future developments.

This work not only covers detection and prevention techniques but also critically evaluates their deployment potential, offering a practical guide for readers. We further extend the discussion by incorporating the potential of \gls{genai}, highlighting its future impact on the field. Resource requirements are classified based on computational complexity and infrastructure demands: minimal refers to lightweight heuristic approaches executable on standard routers with <1GB \gls{ram} and real-time processing capabilities; low encompasses basic \gls{ml}  algorithms (decision trees, simple classifiers) deployable on commodity hardware like Raspberry Pi with 1-4GB \gls{ram} moderate includes standard \gls{ml} approaches (\gls{rf}, \gls{svm}, \gls{nn} requiring dedicated servers with 4-16GB \gls{ram} and sub-second processing latency; high denotes deep learning models and large-scale training requiring \gls{gpu}  acceleration, >16GB \gls{ram}, and cloud computing infrastructure.

%\section{The Realm of Fingerprint Detection} \label{sec: detection}

\section{Taxonomy of IoT Fingerprint Detection} \label{sec: detection}

Fingerprint detection in IoT systems serves a dual purpose, functioning both as a defensive tool and as a potential attack vector. On the defensive side, it enables security monitoring, anomaly detection, access control, and the identification of unauthorized devices. Conversely, adversaries may exploit fingerprinting for reconnaissance, profiling, and targeted attacks that infer device types, identify vulnerabilities, or uncover user behavior patterns.

We categorize detection techniques into three groups, applicable to both defensive and offensive use cases: 
(i) identifying and classifying IoT devices (\S\ref{sec: discovering}), which provides fundamental knowledge about network composition; 
(ii) inferring device activity patterns and user behaviors (\S\ref{sec: infer}), which reveals operational states and usage patterns; and 
(iii) enforcing behavioral policies to detect deviations from expected usage and mitigate risks (\S\ref{sec: policies}), which maintains security and privacy boundaries.

This classification is distinct from the four-group structure used for the paper selection methodology (Section~\ref{sec:methodology}), which served to define search string criteria rather than detection objectives.

%We categorize detection techniques into \textcolor{red}{in Section II-A, we said 4 groups or types --- please fix this for consistency if 3 is what we meant in section II-A -- three groups} applicable to both defensive and offensive use cases: (i) identifying and classifying IoT devices (\S\ref{sec: discovering}), which provides fundamental knowledge about network composition; (ii) inferring device activity patterns and user behaviors (\S\ref{sec: infer}), which reveals operational states and usage patterns; and (iii) enforcing behavioral policies to detect deviations from expected usage and mitigate risks (\S\ref{sec: policies}), which maintains security and privacy boundaries.

In examining these approaches, from supervised learning to protocol-specific heuristics, it is important to consider not only their accuracy but also their deployability in real-world smart home environments. Many methods impose non-negligible traffic, computational or memory resource requirements, which can limit applicability on specific Smart Home scenarios. These trade-offs, together with broader system design considerations and protocol dependencies, are summarized in the take-away boxes below and detailed in the accompanying tables.

\subsection{Discovering IoT Devices in a Network} \label{sec: discovering}

Identifying devices within a network is fundamental for ensuring robust security, efficient network management, and the protection of user privacy. As smart home environments become increasingly saturated with diverse IoT devices, accurate identification becomes essential for monitoring device behavior, detecting unauthorized access, and preventing potential breaches. Effective device discovery also supports better resource allocation and contributes to the overall reliability and performance of home networks.

Device identification refers to the process of determining the specific type and model of IoT devices based on their distinctive network traffic patterns. This is distinct from event or activity detection, which focuses on recognizing specific actions (i.e. turning on a light, adjusting a thermostat, or streaming audio). While identification provides a foundational inventory of networked devices, activity inference adds behavioral context, and together, they enable comprehensive monitoring of IoT ecosystems.

This subsection focuses on detection methods grounded in \textit{passive traffic analysis}, techniques that observe device communication without introducing additional traffic. Passive methods are particularly well-suited for smart home environments, as they preserve normal device operation and avoid disrupting resource-constrained systems. In contrast, alternative methods like Manufacturer Usage Description (MUD) profiles or active probing rely on device compliance or bidirectional communication, which may not be feasible or scalable in heterogeneous home networks.

Traditional identifiers such as MAC addresses are often insufficient for reliable classification due to spoofing, reuse, or lack of granularity \cite{automaticclass}. Similarly, active probing techniques can be intrusive and unsuitable for many IoT devices with limited computational or networking capabilities. These limitations underscore the need for robust passive fingerprinting approaches that are both accurate and unobtrusive. 

Broadly, device discovery techniques fall into three categories based on the types of features they extract from traffic:
\begin{enumerate}[label=(\roman*)]
    \item \textbf{Statistical feature-based methods}, which use quantitative descriptors like packet length distributions, inter-arrival times, and flow durations to identify device-specific patterns;
    
    \item \textbf{Categorical feature-based methods}, which rely on discrete protocol attributes such as IP addresses, port numbers, and \gls{tls} handshake parameters;

    \item \textbf{Hybrid methods}, which combine categorical and statistical features to leverage the strengths of both approaches, often achieving higher accuracy and robustness.
\end{enumerate}

%In the following sections, we review these categories in detail, highlighting representative techniques, their applicability to smart home settings, and their feasibility for real-world deployment.

%To better highlight methodological distinctions, we reorganize the summarized tables of IoT device discovery into three groups, following our taxonomy of \emph{statistical}, \emph{categorical}, and \emph{hybrid} feature-based approaches. 
%This separation emphasizes the key differences in how features are extracted and exploited: 
%(i) \textbf{Statistical} approaches rely on statistical characteristics of traffic (e.g., packet length distributions, entropy, or frequency analysis); 
%(ii) \textbf{Categorical} approaches focus on deterministic or rule-based classification using protocol fields, identifiers, or categorical mappings; and 
%(iii) \textbf{Hybrid} approaches combine statistical and categorical features, often with machine learning or deep learning to improve generalization. 

The following three tables (Tables~\ref{tab:statistical_discovering}, \ref{tab:categorical_discovering}, and \ref{tab:hybrid_discovering}) provide a comprehensive summary of the works within each category, maintaining a consistent structure that details data accessibility, targeted protocols, datasets, adopted techniques, and resource requirements. This systematic organization explicitly highlights the trade-offs between classification accuracy and practical deployability across different fingerprinting approaches, facilitating informed method selection based on specific deployment constraints and performance requirements.
%This organization makes explicit the trade-offs between accuracy and deployability, as discussed in the text.

%---------------------------------------------------------------------------------------------------------

\begin{comment}
\begin{table}[hbt!]
\begin{tabular}{ll}
\textbf{Device Type}                                                    & \textbf{\% of Traffic Sent over HTTPS} \\
Game Consoles                                                           & 90-95\%                                \\
Health Wearables                                                        & 80-85\%                                \\
Home Automation                                                         & 90-95\%                                \\
Smart Assistants                                                        & 75-80\%                                \\
Smart Cameras                                                           & 100\%                                  \\
Smart Speakers                                                          & 95-100\%                               \\
Smart Phones                                                          & 85-90\%                               \\      
\end{tabular}
\caption{Percentage of Web Traffic that are encrypted, across different types of IoT devices.}   
\label{tab:https}
\end{table}
\end{comment}
%\textbf{Flow-based Fingerprints. } 

\subsubsection{ Statistical Feature-based Fingerprinting }

\begin{figure}[h]
\includegraphics[width=\columnwidth]{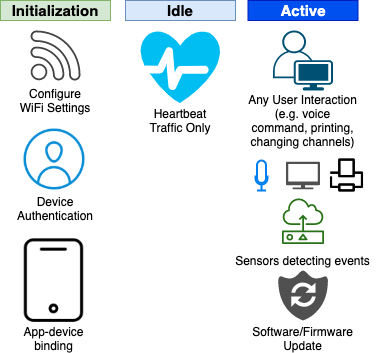}
\caption{An IoT device can be in one of these three operational states. }
%The traffic generated by the IoT device is different depending on its state~\cite{spatial}.}
\label{fig:iotstates}
\end{figure}

A \textbf{traffic flow} refers to a unidirectional sequence of packets sharing the same 5-tuple, source and destination IP addresses, ports, and transport protocol, typically bounded by a timeout or session termination. Even when payloads are encrypted, such flows expose metadata patterns that can be used to characterize device behavior, including packet counts, durations, directionality, and inter-arrival times. While some fingerprinting methods rely on periodic traffic patterns (e.g., DNS queries), flow-based techniques apply more generally to all communications generated by IoT devices, regardless of their operational state (Fig.~\ref{fig:iotstates}). Furthermore, flow-level metadata is relatively costly to conceal or obfuscate, making this approach both practical and resilient.

Existing learning-based solutions in this category can be broadly classified into three groups: traditional machine learning approaches, neural network-based methods, and all other types. IoTSpot \cite{iotspot} is a machine learning-based fingerprinting method trained on real-world labeled traffic from three smart homes. Using \gls{pca} for feature selection and a \gls{rf} classifier, it identifies 11 key flow features (e.g., flow size, duration, average packet size), achieving an F1 score of 0.98 across 21 devices, and 0.92 with just 40 traffic flows. Similarly,\cite{8622243} presents a lightweight flow-based fingerprinting approach targeting six consumer IoT devices (e.g., Nest camera, TP-Link plug), using Random Forests on simple features like packet size and inter-arrival times from the first $N$ packets. 
Different from IoTSpot and~\cite{8622243}, IoTClientDetector~\cite{keepnegotiation} uses temporal patterns in TLS handshakes to detect rogue devices under encrypted traffic. 
Random Forest and \gls{svm} models are also deployed in \cite{9110451} to distinguish IoT from non-IoT devices using just 5 seconds of \gls{http}/\gls{mqtt} traffic. Pinheiro et al. \cite{packetlength}, in their turn, classifies encrypted TLS traffic using packet length statistics within one-second time windows using Random Forest. Msadek et al.~\cite{iotanalysis} compare classifiers (\gls{knn}, \gls{svm}, Extra-Trees, AdaBoost, and \gls{rf}) using features from~\cite{iotspot} and the dataset from~\cite{characterizing}, identifying AdaBoost as the most effective one. Marchal et al.~\cite{audi} and Sivanathan et al.~\cite{inferring} consider unsupervised ML for identification of previously unseen or unauthorized devices, where the latter one also deployed the detection system in a \gls{sdn} controller for adapting training ability.   
%mitigate this by continuous retraining using unsupervised learning within an SDN controller. 

For neural network-based models, Martin et al.\cite{unauthorized} build upon the IoTSpot framework\cite{iotspot} by combining \glspl{cnn} and \glspl{rnn} to improve classification accuracy. Fan et al.\cite{overview_semi,semi-supervised} propose a semi-supervised, \gls{cnn}-based multitask learning approach that reduces reliance on labeled data, thereby simplifying the training process. Ma et al.\cite{spatial} introduce a spatial-temporal fingerprinting technique that leverages \glspl{cnn} and burst distance metrics to estimate the number of devices behind a NAT, while~\cite{9548663} extends this approach with a \gls{nat}-aware method for more accurate multi-device quantification.

In a different direction, Apthorpe et al.\cite{spying_soho} incorporate DNS header information into a neural network classifier to infer both device identity and user activity. Charyyev and Gunes\cite{LSIF} utilize Nilsimsa hashing to compare flows against known fingerprints, enabling flow classification based on similarity. Similarly, Smart Recon~\cite{smart_rec} integrates Nilsimsa hashing with a Multi-Layer Perceptron (MLP) to enable real-time, resource-efficient device identification. Other notable neural network models include IoT-GFCN~\cite{gam_iot} and \cite{soho-safe, traffic_behavior}. 
%%%%\cite{identify_iot_ai} uses HTTP responses and a 517K-parameter neural network for classification. IoT-GFCN~\cite{gam_iot} leverages global attention over packet sequences, achieving 98.88\% accuracy. A two-stage ML-DL classifier is introduced by~\cite{traffic_behavior}, while~\cite{soho-safe} demonstrates high accuracy (99.2\% under NAPT, 97.7\% under VPN) using LSTM-based sequence learning.

Numerous studies leverage statistical techniques in conjunction with machine learning, often emphasizing efficient and robust feature selection. For example,\cite{statfeat1} apply statistical tests such as the t-test and Mann-Whitney U to reduce the feature set by 80\%, with only a 2\% drop in classification accuracy, making it well-suited for resource-constrained smart home environments as proposed in \cite{LSIF-R,Scottish}. Similarly,\cite{statfeat3} focuses on \gls{tcp} inter-arrival times to enable lightweight yet effective device classification. To improve scalability, Haystack\cite{haystack} adopts a flow-sample clustering and classification strategy, achieving 90\% accuracy using just 1\% of total traffic samples. 

Other works\cite{statfeat2,statfeat5} explore multi-layer feature extraction, spanning network, transport, and application layers, to enhance fingerprinting resilience under varying conditions. Hao et al.\cite{IoTTFID} propose an incremental learning model capable of adapting to the introduction of new devices and evolving traffic patterns, even when operating on encrypted traffic, as also addressed in\cite{statfeat4,retraining}. Meanwhile, Chaudhary et al.~\cite{iotperimeter} focus on identifying and extracting the most informative features to improve device classification effectiveness. Sivanathan et al.~\cite{behavior,smartcitiesandcampuses} focus on robustness and retrain classifiers with “white noise” to better distinguish IoT from non-IoT devices.

Several approaches leverage device fingerprinting for security functions such as authentication, authorization, and access control~\cite{bfiot,autonomous,audi,9042946}. Other works focus on adapting fingerprinting techniques to different network environments and their \gls{qos} requirements. For instance, Roy et al.\cite{roy2021automatic} utilize signal-to-noise ratio and modulation-based features for device identification in satellite networks. In contrast, solutions like \cite{statfeat6,autonomous,optimal_feat,progressive} target edge networks, emphasizing lightweight methods suitable for resource-constrained environments. Additionally, Li et al.~\cite{camhunter} explore device-specific fingerprinting by detecting smart cameras through analysis of their periodic heartbeat messages.

\begin{table*}[!h]
\centering
\caption{Summary of statistical feature-based IoT device discovering.}
\begin{tabular}{|>{\centering\arraybackslash}m{1cm}|>{\centering\arraybackslash}m{3cm}|>{\centering\arraybackslash}m{2cm}|>{\centering\arraybackslash}m{4cm}|>{\centering\arraybackslash}m{3cm}|>{\centering\arraybackslash}m{2cm}|}
\hline
\textbf{Work} & \textbf{Data Accessibility} & \textbf{Target Protocol} & \textbf{Dataset} & \textbf{Adopted Technique Keywords} & \textbf{Resources Needed} \\ \hline
\cite{iotspot} & \gls{ap} & Protocol agnostic & Lab, 19 devices & RF, PCA & Low-Moderate \\ \hline
\cite{8622243} & AP, outbound & HTTP, HTTPS & Lab, 4 devices & RF, DT & Moderate \\ \hline
\cite{keepnegotiation} & AP, outbound & TLS Handshake & Lab, 23 devices, Raspberry Pi acts as IoT devices & Rule-based statistical classifier & Moderate \\ \hline
\cite{LSIF} & AP & TCP/UDP & Lab, 23 devices & Locality-sensitive hashing & Minimal \\ \hline
\cite{optimal_feat} & AP, inbound & DNS, mDNS, TLS, HTTP, SSDP, QUIC, MQTT, STUN, NTP, TCP, BOOTP & Public \cite{datasetNew2}, 16 Devices & PCA, SVD, Mutual Information & Minimal \\ \hline
\cite{radtec} & AP, inbound & Protocol agnostic & UNSW dataset \cite{UNSW}, 15 devices & RF, SVM, GBC, GNB & Low \\ \hline
\cite{Scottish} & AP, inbound & Protocol agnostic & UNSW dataset \cite{UNSW} & CatBoost & Moderate \\ \hline
\cite{smart_rec} & AP & Protocol agnostic & Public \cite{LSIF}, 22 devices & RF, GB, KNN, DT & Moderate \\ \hline
\cite{spying} & AP & DNS queries & Lab, 7 Devices & Longitudinal analysis & Minimal \\ \hline
\cite{retraining} & AP, inbound, during setup phase & DHCP, ARP, DNS, TCP/IP & Lab, 43 devices & RF, DT, MLP, GRU, CNN & Minimal \\ \hline
\cite{gam_iot} & Router/AP & Protocol agnostic & UNSW \cite{UNSW} dataset & Global Attention Mechanism, Fully Convolutional Network & Moderate \\ \hline
\cite{traffic_behavior} & Router/AP Passive AP traces & Protocol agnostic & UNSW \cite{UNSW} mixed with lab collected dataset & LSTM, CNN & Moderate \\ \hline
\cite{transformer-abnormal} & AP/Router & Protocol agnostic & N-BaIoT \cite{meidan2018nbaiot} dataset & Transformer models & Moderate \\ \hline
\cite{soho-safe} & AP, both inbound and outbound, VPN enabled & Protocol agnostic & Lab, 14 devices & Packet-level LSTM-RNN, bidirectional LSTM & Moderate \\ \hline
\cite{behaviour_new3} & Both Sniffed OTA and Outbounded AP/Router & Wi-Fi and TCP/IP headers & Two Public dataset \cite{behaviour_new21}, UNSW \cite{UNSW} & Full packet-size distribution with Cosine distance & High \\ \hline
\cite{statfeat1} & SDN-controlled gateway & TLS/SSL & IoT traffic, encrypted flows & LSTM, sequence learning & Moderate \\ \hline
\cite{statfeat3} & ISP & NetFlow data & Real-world traces, ISP-level & Feature selection RF & Scalable \\ \hline
\cite{statfeat2} & AP, outbound & TCP-focused & Weekly IoT traffic, continuous & Feature-based RF & Moderate \\ \hline
\cite{statfeat5} & Router/AP & Protocol agnostic & IoT traces, OpenWrt setup & OpenWrt + RF & Moderate \\ \hline
\cite{statfeat4} & Router/AP & Multi-protocol & Lab, 9 IoT devices & KNN clustering & Moderate \\ \hline
\cite{roy2021automatic} & AP, inbound & TLS & IoT traffic traces, encrypted & t-SNE dimensionality & Moderate \\ \hline
\cite{statfeat6} & SDN controller & Protocol agnostic & Simulated IoT setup & ML with RF & Complex \\ \hline
\cite{autonomous} & AP & Protocol agnostic & IoT traces, real deployment & RF classifiers & Moderate \\ \hline
\cite{bfiot} & Router/AP & Protocol agnostic & Lab, 20 IoT devices & Decision Tree RF & Low \\ \hline
\cite{smartenv} & ISP & DNS & IoT DNS traffic, smart environment & DNS statistical models & Low \\ \hline
\cite{9042946} & SDN-controlled gateway & Protocol agnostic & IoT trace files, mixed protocols & Multi-stage RF+SVM & Moderate \\ \hline
\cite{10154276} & AP, VPN enabled & TLS, HTTP & Simulated IoT setup, encrypted & VPN ML analysis & Moderate \\ \hline
\cite{unauthorized} & AP, inbound & TLS & UNSW-IoT dataset \cite{UNSW} & Supervised RF+SVM & Moderate \\ \hline
\end{tabular}
\label{tab:statistical_discovering}
\end{table*}

% Page break \clearpage

\begin{table*}[!h]
\centering
\caption{Summary of categorical feature-based IoT device discovering techniques.}
\begin{tabular}{|>{\centering\arraybackslash}m{1cm}|>{\centering\arraybackslash}m{3cm}|>{\centering\arraybackslash}m{2cm}|>{\centering\arraybackslash}m{4cm}|>{\centering\arraybackslash}m{3cm}|>{\centering\arraybackslash}m{2cm}|}
\hline
\textbf{Work} & \textbf{Data Accessibility} & \textbf{Target Protocol} & \textbf{Dataset} & \textbf{Adopted Technique Keywords} & \textbf{Resources Needed} \\ \hline
\cite{9688598} & AP, inbound & LAN protocols (e.g., mDNS, SSDP) & 2 Public \cite{smartenv} \cite{smartcitiesandcampuses}, 6 Device types & RF, NB & Moderate \\ \hline
\cite{DevTag2023} & AP, outbound & Multiple Application level Protocols (e.g., TLS, HTTP, SMTP) & Lab, 4 Devices & Rule-based, Learning-Based & Low \\ \hline
\cite{9236963} & AP, outbound & TLS Handshake & Public, 5 Devices & RF, DT, NB, SVM, KNN & Moderate \\ \hline
\cite{disc_iot_enc} & AP, inbound & TLS & UNSW Dataset \cite{UNSW} & TF-IDF (categorize) & Low \\ \hline
\cite{camhunter} & AP, outbound & DNS, TLS & 7 Cameras & RF, KNN, SVM & Moderate \\ \hline
\cite{packetlength} & AP, inbound & TCP/IP header & UNSW Dataset \cite{UNSW} mixed with Lab Devices & KNN, SVM & Moderate \\ \hline
\cite{inferring} & SDN-controlled gateway & Telemetry protocols & Public \cite{datasetNew1}, 10 devices & Clustering-based unsupervised & Moderate \\ \hline
\cite{behaviour_new10} & AP, outbound & TCP/IP header & UNSW Dataset \cite{UNSW} mixed with Lab Devices & CNN, Spatial-temporal analysis, MLP & Moderate \\ \hline
\cite{9110451} & AP, inbound & L3–L5 (multi-protocol) & UNSW Dataset \cite{UNSW} & RF, SVM & Moderate \\ \hline
\cite{iotfinder} & ISP & DNS & Simulated, 53 devices & TF-IDF similarity & Moderate \\ \hline
\cite{iotperimeter} & AP, inbound & TCP/UDP & 2 Public \cite{UNSW} \cite{datasetNew1} & Gaussian Naïve Bayes & Moderate \\ \hline
\cite{iotsentinel} & AP and SDN controller, setup data & L3–L5 (multi-protocol) & Lab, 6 Devices & RF & Moderate \\ \hline
\cite{ziot} & ZigBee/Z-Wave Hub & ZigBee & Lab, 39 devices & Bayes Nets, RF & Minimal \\ \hline
\cite{profiliot} & AP, outbound & Protocol agnostic & Lab, 9 Devices & RF, XGBoost, GBM, Multi-session refinement & Minimal \\ \hline
\cite{profiling-based} & AP, inbound & Protocol agnostic & Public \cite{datasetNew3}, 54 sensor-related devices & Multi-stage ML models & Moderate \\ \hline
\cite{progressive} & SDN-gateway & Protocol agnostic & Not specified & DPI, rule-based matches & Moderate \\ \hline
\cite{IoTTFID} & AP, inbound & Protocol agnostic & UNSW-IoT \cite{UNSW}, 20 devices; Yourthings-IoT \cite{IoTFinder2020}, 30 devices & Transformer, Incremental learning & Moderate \\ \hline
\cite{spying_soho} & AP, outbound & Protocol agnostic & Commercial IoT traffic & 3-NN & Low \\ \hline
\cite{9206046} & ZigBee/Z-Wave Hub & BLE & IoT BLE traffic, proximity & Supervised & Low \\ \hline
\cite{9105052} & Router/AP & DNS & IoT DNS traffic, domain analysis & DNS flow analysis & Moderate \\ \hline
\end{tabular}
\label{tab:categorical_discovering}
\end{table*}

\begin{table*}[!h]
\centering
\caption{Summary of hybrid feature-based IoT device discovering.}
\begin{tabular}{|>{\centering\arraybackslash}m{1cm}|>{\centering\arraybackslash}m{3cm}|>{\centering\arraybackslash}m{2cm}|>{\centering\arraybackslash}m{4cm}|>{\centering\arraybackslash}m{3cm}|>{\centering\arraybackslash}m{2cm}|}
\hline
\textbf{Work} & \textbf{Data Accessibility} & \textbf{Target Protocol} & \textbf{Dataset} & \textbf{Adopted Technique Keywords} & \textbf{Resources Needed} \\ \hline
\cite{smart_rec} & AP & Protocol agnostic & Public \cite{LSIF}, 22 devices & RF, GB, KNN, DT & Moderate \\ \hline
\cite{spying} & AP & DNS queries & Lab, 7 Devices & Longitudinal analysis & Minimal \\ \hline
\cite{retraining} & AP, inbound, during setup phase & DHCP, ARP, DNS, TCP/IP & Lab, 43 devices & RF, DT, MLP, GRU, CNN & Minimal \\ \hline
\cite{gam_iot} & Router/AP & Protocol agnostic & UNSW \cite{UNSW} dataset & Global Attention Mechanism, Fully Convolutional Network & Moderate \\ \hline
\cite{traffic_behavior} & Router/AP, passive AP traces & Protocol agnostic & UNSW \cite{UNSW} mixed with lab dataset & LSTM, CNN & Moderate \\ \hline
\cite{transformer-abnormal} & AP/Router & Protocol agnostic & N-BaIoT \cite{meidan2018nbaiot} dataset & Transformer models & Moderate \\ \hline
\cite{soho-safe} & AP, inbound/outbound, VPN enabled & Protocol agnostic & Lab, 14 devices & Packet-level LSTM-RNN, bidirectional LSTM & Moderate \\ \hline
\cite{behaviour_new3} & Sniffed OTA + outbound AP/Router & Wi-Fi and TCP/IP headers & Public \cite{behaviour_new21}, UNSW \cite{UNSW} & Full packet-size distribution + Cosine distance & High \\ \hline
\cite{statfeat1} & SDN-controlled gateway & TLS/SSL & IoT traffic, encrypted flows & LSTM, sequence learning & Moderate \\ \hline
\cite{haystack} & ISP, inbound/outbound & Protocol agnostic & Public, millions of devices \cite{IoTFinder2020} & Ensemble RF & Scalable \\ \hline
\cite{statfeat3} & ISP & NetFlow data & Real-world traces, ISP-level & Feature selection RF & Scalable \\ \hline
\cite{statfeat2} & AP, outbound & TCP-focused & Weekly IoT traffic, continuous & Feature-based RF & Moderate \\ \hline
\cite{statfeat5} & Router/AP & Protocol agnostic & IoT traces, OpenWrt setup & OpenWrt + RF & Moderate \\ \hline
\cite{statfeat4} & Router/AP & Multi-protocol & Lab, 9 IoT devices & KNN clustering & Moderate \\ \hline
\cite{roy2021automatic} & AP, inbound & TLS & IoT traffic traces, encrypted & t-SNE dimensionality & Moderate \\ \hline
\cite{statfeat6} & SDN controller & Protocol agnostic & Simulated IoT setup & ML with RF & Complex \\ \hline
\cite{autonomous} & AP & Protocol agnostic & IoT traces, real deployment & RF classifiers & Moderate \\ \hline
\cite{bfiot} & Router/AP & Protocol agnostic & Lab, 20 IoT devices & Decision Tree RF & Low \\ \hline
\cite{smartenv} & ISP & DNS & IoT DNS traffic, smart environment & DNS statistical models & Low \\ \hline
\cite{9042946} & SDN-controlled gateway & Protocol agnostic & IoT trace files, mixed protocols & Multi-stage RF+SVM & Moderate \\ \hline
\cite{10154276} & AP, VPN enabled & TLS, HTTP & Simulated IoT setup, encrypted & VPN ML analysis & Moderate \\ \hline
\cite{unauthorized} & AP, inbound & TLS & UNSW-IoT dataset \cite{UNSW} & Supervised RF+SVM & Moderate \\ \hline
\cite{hybrid_edu5} & AP, inbound/outbound & HTTP, HTTPS, DNS & Public, 1,237 devices (66 types), 200+ homes \cite{UNSW} & Statistical analysis & Moderate \\ \hline
\cite{iotanalysis} & Router/AP, setup data & TLS, HTTPS & Lab, 20 devices (5 categories) \cite{UNSW} & RF, k-NN & Moderate \\ \hline
\cite{audi} & Router/AP & TCP, UDP, SSDP, MQTT & Lab, 33 devices (plugs, bulbs, cameras) & k-Means, DBSCAN & Low \\ \hline
\cite{semi-supervised} & Router/AP & TCP, UDP, TLS & Lab, 50 devices (30 IoT, 20 non-IoT), 10\% labeled & CNN, multi-task learning & High \\ \hline
\cite{spatial} & ISP, outbound & DHCP, DNS, TLS & Public, 100+ households, 30 days & DBSCAN, RF & Moderate \\ \hline
\cite{behavior} & Router/AP & MQTT, CoAP, HTTP, TLS & Lab, 28 devices (cameras, lights), 4 weeks \cite{UNSW} & DT, Autoencoders & Moderate \\ \hline
\cite{hybrid_edu1} & AP, inbound & UPnP, SSDP, HTTP, Telnet & Public, 83M devices, 16M households & Statistical analysis & High \\ \hline
\cite{hybrid_edu3} & Router/AP & TCP, UDP, TLS & Lab, 50 devices (30 IoT, 20 non-IoT) & CNN, multi-task learning & High \\ \hline
\cite{hybrid_edu4} & Router/AP & MQTT, HTTP & Real-time, 6 gateways, 6 sensors & Profiling models & Moderate \\ \hline
\cite{hybrid_edu6} & Router/AP & HTTP, TLS, SSDP & Lab, 25 devices (cameras, hubs) & Ontology-based mapping & High \\ \hline
\cite{hybrid_edu7} & Router/AP & HTTP, HTTPS, DNS, TCP, UDP & Simulated, 20 devices (10 IoT, 10 non-IoT) & Flow-level analysis & Moderate \\ \hline
\cite{hybrid_edu8} & Router/AP & HTTP, HTTPS, MQTT, CoAP & Lab, 50 devices (10 categories) & RF, SVM, k-NN & Moderate \\ \hline
\cite{poiriot2022} & Edge switch, inbound/outbound & Protocol agnostic & Public \cite{pingpong}, 14 devices & FSM, DFA, P4 programming & Low \\ \hline
\cite{poiriot2024} & Edge switch, inbound/outbound & Protocol agnostic & Public \cite{pingpong}, 14 devices & FSM partitioning, Device tagging, P4 programming & Moderate \\ \hline
\end{tabular}
\label{tab:hybrid_discovering}
\end{table*}

\subsubsection{Fingerprinting based on Categorical Features}

Categorical features are extracted from raw packet within traffic flows, including elements such as IP addresses, TCP/UDP ports, and protocol-specific identifiers. By analyzing fields in packet headers or payloads, these methods can identify distinctive patterns that serve as reliable signatures of IoT devices. Protocols like TLS, for instance, expose metadata that remain relatively stable across sessions, enhancing the consistency of device identification. Unlike statistical features, categorical attributes are unaffected by packet size or timing, making them particularly effective for fingerprinting devices, even under encrypted traffic conditions. However, their effectiveness can diminish in scenarios where protocols or port numbers are dynamically modified. Notably, few existing fingerprinting approaches have fully leveraged categorical feature-based detection, suggesting a promising avenue for further research.

A couple of schemes \cite{iotfinder,9688598} focus mainly on analyzing protocol data, e.g., DNS queries, to fingerprint IoT devices. For instance, IoTFinder \cite{iotfinder} combines the number of communicated domains and the frequency of such communications of IoT devices for effective fingerprinting, as IoT devices differ significantly in such metrics. Similarly, Roemsri et al. \cite{9688598} exploit traffic from DNS, mDNS, NUPNP, and SSDP protocols and measure Levenshtein distance between network traffic from various devices for fingerprinting.  
Valdez et al. \cite{disc_iot_enc}, in their turn, use TLS handshake as one of the main features to distinguish IoT devices. They first measure the frequency of each features in TLS sessions across various devices and their types to assign weights to the chosen features to uniquely identify devices.

%\textcolor{red}{this one seems hybrid; am I missing something? --  } \textcolor{blue}{--Right, I'll move it to the hybrid.}

\subsubsection{Fingerprinting based on Hybrid Features}

Hybrid feature-based fingerprinting combines both statistical and categorical features to improve the accuracy and reliability of IoT device identification. This approach leverages statistical data, such as packet size and timing, along with categorical information from packet headers, providing a multidimensional view of traffic patterns. By integrating these feature types, hybrid methods can overcome the limitations of individual approaches, improving detection rates under varying network conditions. These methods are particularly effective when traffic includes identifiable header fields and consistent behavioral patterns, although increased computational complexity may present challenges in resource-limited environments \cite{hybrid_edu1}.

Most of the work in this space consists of machine learning-based approaches to IoT device classification. For instance, ProfilIoT \cite{profiliot} employs a multi-stage classification pipeline that first distinguishes IoT from non-IoT traffic using categorical features such as HTTP headers, before performing device-specific classification. A similar hierarchical approach is adopted in \cite{hybrid_edu1}, which utilizes both statistical features (e.g., traffic volume, flow duration) and categorical features (e.g., destination IPs, port numbers) drawn from network and application layer traffic. Z-IoT \cite{ziot}, MIB-IoT \cite{profiling-based}, and PoirIoT \cite{poiriot2022,poiriot2024} also follow machine learning-based designs, with Z-IoT integrating protocol-specific categorical attributes from ZigBee and Z-Wave traffic alongside statistical features, and PoirIoT using packet metadata with finite state machines for real-time processing. In \cite{9236963}, Nguyen-An et al. first compute the information entropy of various traffic attributes (e.g., IP addresses, port numbers, and packet sizes) to capture behavioral variability, then apply a Random Forest classifier to achieve accurate device instance classification. IoTSentinel \cite{iotsentinel} also follows a hybrid machine learning approach, combining statistical features (e.g., packet lengths, flow durations) and categorical attributes extracted from header fields (e.g., destination IP, TCP options, IP padding) to fingerprint devices. Uniquely, it integrates this fingerprinting with a security mechanism for automatic device-specific access control. 

Semi-supervised classifiers are introduced in \cite{semi-supervised,hybrid_edu3}, further demonstrating the versatility of learning-based models in this domain. Specifically, a neural network based model is proposed in \cite{semi-supervised}. IoTminer \cite{hybrid_edu6} develops an \gls{lstm}-based classifier by incorporating semantic extraction from packet payloads along with statistical features. 
Traffic characterization is the key focus in \cite{hybrid_edu5,hybrid_edu7}. Finally, the authors of \cite{hybrid_edu4,hybrid_edu8,DevTag2023} compare various ML models to assess their effectiveness in the detection of IoT devices using both statistical and categorical features. For example, DevTag \cite{DevTag2023} benchmarks IoT device fingerprinting by integrating both rule-based (nonML-based) and model-based classifiers. Their results show that the model-based approach has significant advantages in distinguishing coarse-grained information like device type and vendor. However, such approaches may struggle detecting product information due to the large number of labels, where rule-based approaches are effective.  

%IoTSentinel is a security system developed by Miettinen et al. that performs device identification and automatic device-specific access control \cite{iotsentinel}. IoTSentinel uses an ML-based classifier that relies on header data (e.g. Destination IP address, TCP options, IP padding) as well as packet lengths and flow duration to fingerprint and monitor active devices on a network. Furthermore, IoTSentinel is deployed on an SDN controller that blocks traffic from unauthorized or potentially vulnerable IoT devices.

\findingsbox{
Statistical, categorical, and hybrid feature-based fingerprinting offer complementary strategies for IoT device identification, each with distinct advantages, limitations, and application scenarios (See Tables~\ref{tab:statistical_discovering},~\ref{tab:categorical_discovering},~\ref{tab:hybrid_discovering}). As such, the choice of approach should align with the specific demands of the deployment context.
Statistical methods are well-suited for lightweight, resource-constrained environments due to their efficiency and simplicity. In contrast, categorical approaches are more effective in scenarios involving encrypted traffic or heightened privacy concerns, as they are less reliant on packet size or timing. Hybrid methods aim to combine the strengths of both, but fully realizing their potential will require advanced integration techniques, e.g., attention mechanisms or semantic feature modeling, to improve cross-domain learning.
To ensure ethical and scalable device identification, especially in privacy-sensitive settings, the development of privacy-preserving frameworks like federated learning will be essential. These techniques can enable secure model training without compromising sensitive data, supporting responsible innovation in IoT fingerprinting. }

\subsection{Inferring Events} \label{sec: infer}

Inferring events from IoT network traffic involves detecting device states, transitions, and user behaviors using passive traffic analysis. These techniques rely on extracting both statistical features (e.g., packet size, inter-arrival time, traffic volume) and categorical features (e.g., packet direction, protocol type, device ID) from network traces. In combination with machine learning models and temporal correlations, these features enable the reconstruction of meaningful events from raw traffic.

While event inference techniques can improve automation and security, they also pose significant privacy risks. Adversaries can deduce device states (e.g., ‘Is the smart TV on?’), transitions (e.g., ‘Did the smart lock activate?’), and even content-level information (e.g., ‘What song is streaming?’). Beyond device-level inference, attackers can profile user activities (e.g., ‘Is the user watching TV?’ or ‘Is the user at home?’) by analyzing traffic patterns in both encrypted and unencrypted communications. Threat actors, including ISPs, surveillance agencies, and eavesdroppers, can correlate flow metadata with behavioral patterns, making privacy protection a critical concern. Existing work in this domain generally falls into two categories: \textit{detecting device state transitions} and \textit{classifying discrete events and user activities}.

\subsubsection{Identifying Device State Transitions}

Identifying device state transitions involves detecting or predicting changes in an IoT device's operational status (i.e. switching from off to on, entering standby, or transitioning between active and idle modes). The scope of existing work in this area varies depending on the specific states being targeted and the underlying traffic analysis methods employed. Table~\ref{tab:iot_state_transitions} provides a comprehensive overview of the techniques and datasets used in this domain.

\begin{table*} [!h]
\caption{Summary of device state transition detection.}
\label{tab:iot_state_transitions}
\begin{tabular}{|>{\centering\arraybackslash}m{1cm}|>{\centering\arraybackslash}m{3cm}|>{\centering\arraybackslash}m{2cm}|>{\centering\arraybackslash}m{4cm}|>{\centering\arraybackslash}m{3cm}|>{\centering\arraybackslash}m{2cm}|}
\hline
\textbf{Work} & \textbf{Accessibility of the Data} & \textbf{Target Protocol} & \textbf{Dataset} & \textbf{Adopted Technique} & \textbf{Resources Needed} \\ \hline

\cite{pingpong} & Sniff both OTA and AP/Router & General to both Wi-Fi, Zigbee, TCP/IP headers & Mixed, Lab collected 19 devices, MonIoTr dataset \cite{behaviour_new13}, 21 devices & Flow signature-based state transition detection & Low \\ \hline

\cite{behaviour_new5} & IoT behavior inference in smart home & TCP/IP headers & Two Public Datasets \cite{pingpong} \cite{behaviour_new13} & Flow signature-based state transition detection & Low \\ \hline

\cite{behaviot} & Real-world Smart-Home Device & L3 and above headers & Lab, 50 devices & DFT, DBScan, Probabilistic state machine & Moderate. \\ \hline

\cite{behaviour_new14} & IoT event classification & TCP/IP headers & MonIoTr Dataset \cite{behaviour_new13} & Multiple ML such as K-NN, DT, RF & Moderate \\ \hline

\cite{behaviour_new16} & AP, Inbound & TCP/IP headers & Mixed, Lab collected 16 devices, MonIoTr dataset \cite{behaviour_new13}, 29 devices & White-box signature-based method & Low \\ \hline

\cite{behaviour_new20} & Sniffed OTA, then decrypt with WPA key & SSL/TLS, HTTP, DNS, NTP, and WEAVE & Lab, 2 Nest devices & Correlated periodicity and burst sizes with statistical and temporal analysis & Low \\ \hline

\cite{behaviour_new13} & AP/Router, both Inbound and Outbound & L3–L5 (multi-protocol) & 81 Lab Devices & SVM, RF & Low \\ \hline

\end{tabular}
\end{table*}

For example, PingPong \cite{pingpong} detects state transitions (e.g., ON/OFF) by analyzing structured packet exchanges, like MQTT or HTTP traffic between IoT devices and cloud backends. The method is lightweight, relying solely on packet length and direction. Carson et al. \cite{Commodity} effectively “SDN-izes” PingPong, a controller mines PingPong-style which generates signatures based on traffic direction and the length, then installs them via P4Runtime as match-action rules, while an Intel Tofino switch executes the finite-state matcher entirely in its data plane, giving real-time, line-rate fingerprinting with no traffic mirroring, sustaining 100\% device detection and about 70\% event detection at 40 Gbps. By adding FSM partitioning and per-device tagging, the authors evolve their solution \cite{Commodity} prototype into PoirIoT \cite{PoirIoT}, raising event-detection from to 92\% at 40 Gbps. DESEND \cite{behaviour_new5} builds on PingPong by removing the dependency on ordered packet pairs, instead using unordered packet size signatures to enable faster and more robust state transition detection.

BehavIoT \cite{behaviot} develops a multi-stage classify IoT states system based on the observation that most IoT device behaviors are either periodic (e.g., heartbeats and status updates) or event-driven (e.g., user interactions). The authors deploy Discrete Fourier Transform (DFT) and Autocorrelation-based spectral analysis to seperate the heartbeat message (e.g., high frequency) from event traffic and perform supervised event classifications.  BehavIoT can identify deviations such as device relocation, network outage, and misconfiguration from normal activities. While BehavIoT develops multi-stage state transition analysis, the authors in \cite{behaviour_new14} assess the impact of multiple ML models on state transition traffic. specifically, they examine how different interaction modes (e.g., voice assistants, mobile apps, or direct device control) impact classification accuracy.
IoTAthena \cite{behaviour_new16} unveils IoT device activities transitioning by analyzing structured sequences of IP packets with inter-packet time intervals based on two polynomial-time algorithms sigMatch and actExtract.  Finally, \cite{behaviour_new20} demonstrates how Nest Thermostat transitions between Home and Auto-Away modes.

\subsubsection{Classifying Events and Profiling User Activities}

This category includes techniques aimed at classifying discrete IoT-triggered events (e.g., door openings, voice commands, sensor activations) and inferring higher-level user activities (e.g., watching TV, arriving home) based on patterns in network traffic. While event classification focuses on detecting and labeling short-lived, device-specific occurrences, user activity profiling aggregates and interprets these events to reconstruct human behavior over time. Both rely on traffic metadata for example packet size, timing, and direction, and often employ machine learning or statistical models to infer activity, even in the presence of encryption or obfuscation. Table~\ref{tab:iot_user_activities} summarizes the key approaches and their characteristics in this area.

\begin{table*} [!h]
\caption{Summary of event classification and user activity profiling.}
\label{tab:iot_user_activities}
\begin{tabular}{|>{\centering\arraybackslash}m{1cm}|>{\centering\arraybackslash}m{3cm}|>{\centering\arraybackslash}m{2cm}|>{\centering\arraybackslash}m{4cm}|>{\centering\arraybackslash}m{3cm}|>{\centering\arraybackslash}m{2cm}|}
\hline
\textbf{Work} & \textbf{Accessibility of the Data} & \textbf{Target Protocol} & \textbf{Dataset} & \textbf{Adopted Technique} & \textbf{Resources Needed} \\ \hline

\cite{behaviour_new12} & Real-world IoT event traces & Protocol agnostic & Lab collected dataset (11 devices) mixed with generated synthetic data & E2AP, edit distance minimization, Theoretical proof & Moderate \\ \hline

\cite{behaviour_new11} & TLS-tunnel building upon UDP & TLS, UDP & UNSW \cite{UNSW} & MLP & Moderate \\ \hline

\cite{behaviour_new9} & ZigBee/Z-Wave traffic monitoring & ZigBee, Z-Wave & CICIoT2022 dataset \cite{CICIoT22} & sequence models (e.g., BiLSTM) & High \\ \hline

\cite{peekaboo} & Sniff OTA & WiFi, ZigBee, BLE & 22 commercial IoT devices, multi-protocol dataset & Hidden Markov Model (HMM), supervised ML & Moderate\\ \hline

\cite{homesnitch} & Real-world deployment in a smart home & TCP/IP headers & YourThings \cite{alrawi2019sok} Dataset, 46 devices & Semantic behavior classification, supervised ML & Moderate \\ \hline

\cite{behaviour_new22} & Packet-length analysis in bidirectional traffic & TCP/IP headers & Two Public Datasets \cite{pingpong} \cite{behaviour_new13}  & Unsupervised signature extraction via hierarchical clustering of packet bursts & Moderate \\ \hline

\cite{behaviour_new23} & AP, both LAN and WAN & WiFi, Bluetooth, Zigbee & Lab, 19 devices & Traffic pattern clustering with distance calculation & Moderate \\ \hline

\cite{behaviour_new7} & IoT user activity inference & TCP/IP headers & Lab, 19 devices & Wavelet decomposition for feature extraction and Naïve Bayes classification & Low \\ \hline

\cite{behaviour_new18} & IoT security enforcement using behavior analysis & WiFi, Zigbee, Z-Wave, and BLE & Lab, 5 devices, Samsung SmartThings with 183 SmartApps & RF for single event classification, sliding window to generate sequence for detected events, app analysis & Moderate \\ \hline

\cite{behaviour_new19} & Event-sequence-based user activity profiling & DNS, NTP, MQTT, HTTPS, meta and partial payload & Lab, 11 devices & Generate signature based on \cite{behaviour_new16}, then apply approximate signature matching algorithm & Moderate\\ \hline

\cite{behaviour_new6} & Signature-based IoT device activity detection & Encrypted traffic & Multiple datasets across smart home IoT devices & Packet-level dynamic signature extraction, adaptive signature length & Moderate\\ \hline

\cite{behaviour_new24} & Encrypted VPN IoT traffic analysis & TCP/IP headers & Three Public datasets, PINGPONG \cite{pingpong}, UNSW \cite{UNSW}, YourThings \cite{alrawi2019sok} & DBSCAN for signature extraction, stream based state machines for classification & High \\ \hline

\cite{behaviour_new8} & Spatial context analysis of IoT events &  TCP/IP headers & Two Public Datasets \cite{pingpong} \cite{behaviour_new13}  & Similar to \cite{behaviour_new22} & Moderate \\ \hline

\cite{behaviour_new10} & Hidden IoT device event detection behind NAT & TCP/IP header & UNSW Dataset \cite{UNSW} Mixed with Lab Devices& CNN, Spatial-temporal analysis, MLP & Moderate \\ \hline

\cite{behaviour_new17} & Sniffed OTA, WPA encrypted & WiFi headers & Lab, 23 devices & Word2Vec embeddings and Attention-based LSTM for sequence modeling of user's activity & Moderate \\ \hline

\cite{behaviour_new2} & Passive gateway traffic & TLS/SSL & 11 Devices from dataset \cite{behaviour_new19} & Coordinate Descent Optimization & Moderate\\ \hline

\end{tabular}
\end{table*}

Peek-a-Boo \cite{peekaboo} introduces a machine-learning pipeline for classifying device events and user activities from encrypted traffic by analyzing packet size and inter-arrival times. It employs an HMM-based model to identify interactions, achieving over 90\% accuracy. PINBALL \cite{behaviour_new22} refines PingPong by introducing bidirectional packet-length signatures, enhancing resilience against packet loss, retransmissions, and network fluctuations. IoTMosaic \cite{behaviour_new19}, in turn, presents an event-sequence-based classification system that correlates event traffic with user activities. An extension of PINBALL and IoTMosaic is presented in \cite{behaviour_new12}, introducing approximate signature matching to handle missing, unordered, or ambiguous device events. IoTDuet \cite{behaviour_new8} distinguishes between local and remote events by correlating IoT event signatures with control traffic to cloud servers.

Several studies have focused on improving the resource efficiency of detection schemes by leveraging a small set of packet-level statistical features \cite{behaviour_new6,behaviour_new23}. For instance, unlike PingPong, the approach in \cite{behaviour_new6} dynamically determines the optimal number of packets needed for signature generation, enabling accurate detection of both binary and multi-event scenarios while being more resilient to network jitter. Similarly, Yao et al. \cite{behaviour_new23} filter out irrelevant packet-level features and propose an adaptive fingerprinting method tailored for event detection. Chen et al. \cite{behaviour_new9} also retrieve useful features from a small number of ZigBee and Z-Wave traffic packets for classification.

Calda et al. \cite{behaviour_new11} examine the feasibility of inferring device behavior over encrypted Layer 2 TLS tunnels. Their findings show that activities can still be classified despite encryption, leading them to recommend traffic padding as a countermeasure. Ma et al. \cite{behaviour_new10} focus on detecting events behind NATs for improved ISP-level device monitoring.

At a broader level, user activity inference seeks to reconstruct human-centric patterns such as daily routines, location-based behaviors, or interaction habits with smart home appliances. Xue et al. \cite{behaviour_new2} demonstrate the correlation between discrete events and user actions by formulating an Events to Activities (E2A) and Events to Activity Patterns (E2AP) inference system, using constrained matching and unsupervised learning. IoTBeholder \cite{behaviour_new17} successfully infers behaviors by monitoring signal fluctuations and usage cycles in WiFi traffic. Similarly, IoTGaze \cite{behaviour_new18} detects policy violations and unexpected behaviors through event sequence dependencies, offering utility in contexts like elderly care.

Lin et al. \cite{behaviour_new12} explore behavior inference in the presence of missing or out-of-order events due to device malfunctions. TrafficSpy \cite{behaviour_new24} evaluates behavioral inference risks even over VPN-encrypted traffic, using a \gls{fhmm} to disaggregate traffic and expose habits. Engelberg et al. \cite{behaviour_new3} show that full packet-size distributions, beyond simple statistics, allow classification even with VPNs and padding. Xu et al. \cite{behaviour_new7} use wavelet analysis to isolate low-frequency, user-triggered traffic from background noise for accurate inference. Finally, a comparative study in \cite{behaviour_new4} shows that Deep Forest (DF) models outperform 1D-CNNs and LSTMs for behavior classification under various network conditions.

\findingsbox{
Tables~\ref{tab:iot_state_transitions} and~\ref{tab:iot_user_activities} summarize the IoT state transition detection and user activity profiling schemes discussed in this section. The approaches rely on packet-level signatures, statistical analysis, and machine learning models. Despite encryption, metadata remains a key leakage vector, enabling adversarial inference with high accuracy. Among the most effective approaches, packet-sequence signatures have demonstrated robustness in detecting device state transitions with minimal computational overhead. Deep learning-based methods, including \glspl{cnn}, \glspl{lstm}, and \glspl{fhmm}, significantly improve classification accuracy but require extensive training data and higher processing power. Adaptive classification models show promise in handling network variability and missing data while maintaining high precision. Studies also confirm that \gls{vpn} encryption alone does not prevent behavioral inference, necessitating stronger privacy-preserving architectures. Future research should focus on real-time traffic obfuscation, efficient metadata padding techniques, and standardized benchmarks for evaluating inference resistance, ensuring robust privacy protection in evolving IoT environments. In addition, we found that most existing event fingerprinting studies rely on manually labeled data to initiate inference. This approach makes the strong assumption that adversaries possess detailed knowledge of the user's home environment, an assumption that is often unrealistic in real-world scenarios. A promising future direction would be to develop methods capable of learning device events directly from unlabeled environments.}

\subsection{Enforcing Event-Level Policies} \label{sec: policies}

Event-specific IoT fingerprints provide a mechanism to enforce security and privacy policies at runtime, ensuring that device behaviors align with user expectations and regulatory compliance. Various research efforts have demonstrated that IoT devices often leak sensitive data or communicate unexpectedly, posing security risks to users. To mitigate these risks, fingerprinting techniques have been used to monitor network activity, classify behaviors, and enforce security policies on IoT traffic. Several approaches have been proposed, ranging from DNS-based fingerprinting to ML-driven classification, enabling researchers to define event-driven policies that restrict or filter traffic based on device behavior. In the following sections, we present a range of methods and classify them according to their enforcement strategies.

\subsubsection{Network-layer Policy Enforcement}

Network-layer policy enforcement relies on packet-level fingerprinting to distinguish between legitimate and non-essential traffic. These methods often use SDN-based control, DNS filtering, and encrypted traffic analysis. Table~\ref{tab:network_policy} summarizes the main approaches in this category. Policies can be enforced by dropping specific device behaviors, such as blocking motion-triggered uploads from Ring Doorbells. 

For example, HomeSnitch is a ML-based IoT traffic classification framework that enforces policies using SDN-based access control \cite{homesnitch}. The system monitors encrypted traffic patterns, identifying devices and behaviors through statistical fingerprinting of packet size and duration. 

\begin{table*}[t]
\caption{Summary of network-layer policy enforcement techniques.}
\label{tab:network_policy}
\begin{tabular}{|>{\centering\arraybackslash}m{1cm}|>{\centering\arraybackslash}m{3cm}|>{\centering\arraybackslash}m{2cm}|>{\centering\arraybackslash}m{4cm}|>{\centering\arraybackslash}m{3cm}|>{\centering\arraybackslash}m{2cm}|}
\hline
\textbf{Work} & \textbf{Data Accessibility} & \textbf{Target Protocol} & \textbf{Dataset} & \textbf{Adopted Technique Keywords} & \textbf{Resources Needed} \\ \hline

\cite{blocking} & AP, outbound & HTTPS, MQTT & Lab, 31 devices & Policy-based classification, automated testing tools & Moderate \\ \hline
\cite{homesnitch} & AP, inbound & Protocol agnostic & Lab, 20 devices & Supervised ML, semantic classification & Moderate \\ \hline
\cite{policy3} & SDN controller, setup data & OpenFlow & Simulated, 1000 flow rules & Policy translation, prioritization algorithms & Moderate \\ \hline
\cite{policy4} & AP, inbound & EPCglobal RFID & Lab, 30 mobile RFID systems & Mathematical modeling, hash-based verification & Low \\ \hline
\cite{policy5} & AP, inbound & HTTPS, MQTT & Lab, 40 devices & Genetic Algorithm & Moderate \\ \hline
\cite{policy6} & SDN controller, inbound & OpenFlow, IP & Lab, 500 hosts & RF, NB & Moderate \\ \hline
\cite{policy8} & AP, inbound & Blockchain protocols & Lab, 10 devices & Blockchain-based enforcement, cryptographic validation & Moderate \\ \hline
\cite{policy9} & AP, outbound & HTTPS, REST API & Lab, 140 test cases & Heuristic evaluation, automated test generation & Moderate \\ \hline

\end{tabular}
\end{table*}

A similar machine learning and SDN-based access control mechanism is proposed in \cite{policy6}. Rosendo et al. introduce HACFlow, an autonomic, policy-based authorization framework for OpenFlow networks \cite{policy3}. Their framework enables fine-grained access control rules, conflict resolution mechanisms, and dynamic adaptation based on real-time network conditions. Another notable contribution is TBAC, a Tokoin-based access control scheme that integrates blockchain and Trusted Execution Environment (TEE) technologies to ensure policy compliance and auditable enforcement in IoT networks \cite{policy8}. TBAC introduces fine-grained, tokenized access control for IoT devices, preventing unauthorized access and ensuring secure policy enforcement.

A different policy-based approach for IoT access control in smart homes is proposed by Alshaboti et al., where security policies are automatically generated based on user preferences and IoT workflows \cite{policy5}. Their method employs heuristic-based device selection algorithms (including Genetic Algorithms) to enforce security policies at the application layer. Mandalari et al. \cite{blocking} propose a DNS-based fingerprinting approach to classify IoT network traffic into essential and non-essential categories. By programmatically testing \gls{dns} requests, they identify domains that are not critical for device functionality and automatically block them. Finally, Nafis et al. \cite{policy9} developed an automated policy evaluation platform VetIoT and evaluated three popular run-time policy enforcement tools: IoTGuard, ExPAT, and PatrIoT \cite{policy9}. The study revealed that ExPAT and PatrIoT were more consistent in enforcing network-layer security policies than IoTGuard, which struggled with compliance deviations.

\subsubsection{Behavioral Policy Enforcement}

Behavioral policy enforcement focuses on device activity monitoring and anomaly detection to regulate interactions between IoT devices and external entities. Table~\ref{tab:behavioral_policy} provides an overview of the main techniques used in this domain. 

In the already mentioned \cite{peekaboo} the authors also introduce an ML-driven privacy attack that can infer IoT device states and user activities, even when communications are encrypted. LeakyPick \cite{leakypick} identifies smart home devices that secretly record and transmit audio to cloud services. It works by probing the environment with synthetic audio and analyzing network traffic to detect suspicious outbound transmissions. 

\begin{table*}[t]
\caption{Summary of behavioral policy enforcement techniques.}
\label{tab:behavioral_policy}
\begin{tabular}{|>{\centering\arraybackslash}m{1cm}|>{\centering\arraybackslash}m{3cm}|>{\centering\arraybackslash}m{2cm}|>{\centering\arraybackslash}m{4cm}|>{\centering\arraybackslash}m{3cm}|>{\centering\arraybackslash}m{2cm}|}
\hline
\textbf{Work} & \textbf{Data Accessibility} & \textbf{Target Protocol} & \textbf{Dataset} & \textbf{Adopted Technique Keywords} & \textbf{Resources Needed} \\ \hline

\cite{iotwatch} & AP, outbound & HTTPS, MQTT & Public \cite{realtime}, 380 SmartThings apps & NLP, taint-sink classification & Moderate \\ \hline
\cite{leakypick} & AP, outbound & HTTPS, TLS & Lab, 8 devices & Statistical traffic analysis & Low \\ \hline
\cite{voicetraffic} & AP, outbound & HTTPS & Lab, 150,000 traces (Amazon Echo, Google Home) & CNN, LSTM & High \\ \hline
\cite{peekaboo} & AP, inbound & WiFi, ZigBee, BLE & Lab, 22 devices & HMM, supervised ML & Moderate \\ \hline
\cite{policy1} & AP, outbound & MQTT, HTTP & Lab, 20 devices & Semantic nets, corpus-based similarity & Moderate \\ \hline
\cite{policy2} & AP, inbound & HTTP, CoAP & Lab, 50 devices & DT, SVM & Moderate \\ \hline
\cite{policy7} & AP, outbound & HTTPS, TLS & Lab, 11 manufacturers & Packet inspection, policy compliance analysis & Moderate \\ \hline

\end{tabular}
\end{table*}

A couple of schemes focus on real time detection to take a quick measurement. 
IoTWatcH \cite{iotwatch} analyzes IoT application behavior at runtime to detect privacy violations and unauthorized data sharing. By classifying privacy-related events using NLP-based log analysis, it provides users with real-time notifications of potential data leaks. 
Subahi et al. \cite{policy7} offer a compliance monitoring framework for IoT privacy policies. Their system monitors real-time traffic from IoT devices and ensures that it adheres to predefined privacy policies, helping detect policy violations in cloud-based IoT environments.

An adaptive access control model for IoT environments is proposed by Alkhresheh et al. \cite{policy2}. Their method uses machine learning to dynamically refine access policies based on behavioral deviations, preventing unauthorized access and ensuring adaptability to evolving security threats.

\findingsbox{
While ML-based fingerprinting methods have shown great potential to enforce security policies in IoT environments, their robustness against adversarial evasion remains an open challenge \cite{policy1, policy6, policy7}. The effectiveness of approaches such as HomeSnitch \cite{homesnitch} and VetIoT~\cite{policy9} demonstrates the viability of automated policy enforcement, yet their reliance on pre-trained models introduces limitations in real-time adaptability. Furthermore, while blockchain-based policy enforcement frameworks, such as TBAC \cite{policy8}, offer strong auditability and decentralized security, their computational overhead remains a barrier to large-scale adoption. 
Dynamic \gls{acl} policy frameworks~\cite{policy6} and heuristic-driven access control~\cite{policy5} appear to provide the best balance between adaptability and computational efficiency. However, their generalizability across diverse IoT ecosystems has not yet been thoroughly validated. A summary of existing schemes is presented in Tables~\ref{tab:network_policy} and~\ref{tab:behavioral_policy}.

Despite advances in IoT policy enforcement, adversaries can still exploit system vulnerabilities by using fingerprint obfuscation, adversarial ML attacks, and covert data exfiltration. Attackers modify packet structures or inject random noise to evade detection. Machine learning models used for fingerprinting can be tricked with adversarial inputs crafted \cite{policy1}. IoT malware can disguise unauthorized transmissions as legitimate device activity, bypassing conventional filters \cite{policy4}. 

Thus, future work should focus on designing hybrid enforcement approaches that integrate network layer filtering with behavioral anomaly detection to enhance real-time adaptability. Additionally, addressing the adversarial robustness of ML-based fingerprinting models remains critical, particularly through adversarial training and privacy-preserving learning techniques. Lastly, while blockchain-backed enforcement improves transparency, its integration with lightweight cryptographic mechanisms may be necessary to mitigate performance overhead without compromising policy compliance. In addition, we found that much of the policy-related research relies on manually designing rules, as the process is often platform-dependent. For example, while high-level policy designs may be similar across different routers/APs, enforcing those policies typically requires different implementations for each platform. With the rise of \glspl{llm}, we see the potential for automating this intermediary step, translating high-level policy designs into platform-specific rule implementations, on demand and in ad-hoc scenarios. This represents another promising future research direction. }

\section{The Domain of Fingerprinting Prevention} 
\label{sec:prevention}

\begin{figure}[t]
\centering
\includegraphics[scale=0.48]{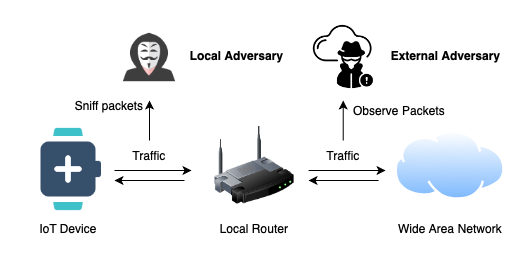}
\caption{An example of local and external adversaries.}
\label{fig:adversaries}
\end{figure}

Fingerprinting of IoT devices (Figure~\ref{fig:adversaries}) enables malicious actors to infer private information about devices and their users \cite{pingpong,privacyguard,homesnitch}. For example, an attacker can fingerprint the traffic patterns of a smart plug to reveal when a user is at home or an appliance is on \cite{audi}. Therefore, developing defenses against privacy attacks based on IoT device fingerprinting is a major concern in academia and industry. This section details the techniques and solutions on defensive countermeasures to protect against unwanted fingerprinting.

\begin{figure}[b]
\centering
\includegraphics[scale=0.9]{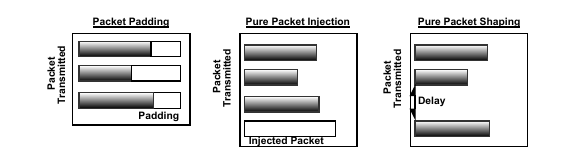}
\caption{Main defense approaches to prevent IoT fingerprinting, including packet padding, traffic injection, and traffic shaping.}
\label{fig:defense_approaches}
\end{figure}

Fingerprinting can be performed based on statistical and categorical features of packet payload or metadata, as discussed in Section~\ref{sec: detection}. Consequently, defending against these techniques requires obfuscating such information, which we categorize into three groups (Figure~\ref{fig:defense_approaches}): (1) packet padding, (2) traffic injection, and (3) traffic shaping. These techniques vary in terms of their obfuscation mechanisms, each focusing on specific types of information leakage.

\textbf{Threat Models:} There are two types of adversaries considered: \textit{local} and \textit{external} (Figure~\ref{fig:adversaries}). Local adversaries are malicious actors who can observe IoT traffic from within a local network. Traffic sniffers and WiFi eavesdroppers are the most common local adversaries. External adversaries are malicious actors who can only observe IoT traffic once it leaves the local network. Malicious routers and ISPs are examples of external adversaries \cite{smartshaping}. The threat model assumes these adversaries use fingerprints to infer information about IoT devices, their types, and events, which can leak sensitive or private information about users. While some contributions consider both local and external adversaries in their threat models, others mostly consider external adversaries.

%\textbf{Threat Models:} There are two types of adversaries considered: \textit{local} and \textit{external} (Figure \ref{fig:adversaries}). Local adversaries are malicious actors who can observe IoT traffic from within a local network. Traffic sniffers and WiFi eavesdroppers are the most common local adversaries. External adversaries are malicious actors who can only observe IoT traffic once they have left the network. Malicious routers and ISPs are examples of external adversaries \cite{smartshaping}. The threat model assumes these adversaries have fingerprints that can infer information about IoT devices, their types, and events, which can leak sensitive or private information about users. While some contributions consider both local and external adversaries in their threat models, others mostly consider external adversaries. 
%I tried to define the general threat model for all prevention schemes
%then, in each prevention scheme, I tried to define the most common threat model

% \textcolor{red}{ add a figure for the threat model.} figure_adversary

\subsection{Packet Padding} \label{sec: padding} 

Packet padding refers to adding dummy bytes to some or all of the outgoing packets (Figure \ref{fig:defense_approaches}). Packet padding can be implemented at the link, network, transport, or application layer. For example, Ethernet packets can be padded to reach the minimum transmission size. Application layer padding is a successful protection scheme against website fingerprinting \cite{app-level-padding}. There are a variety of methods of padding packets. For example, one can pad all packets with the same number of bytes, which may fail to obfuscate traffic size metadata as the probability distribution of padded packet sizes largely remains the same as that of the original packets. A popular approach is to pick a random size between the original packet size and the Maximum Transmission Unit (MTU) \cite{nonadaptive,randommtu,randommtu2}. However, in this approach, original small-size packets tend to stay small while larger packets only get larger; thus, they may not be effective against length-based fingerprinting. One can also pad all packets to the MTU to get the best possible obfuscation, which comes at the price of high bandwidth usage \cite{defending-side-channel}. In the following, we present works that focus on obfuscation, bandwidth usage, or balance both. These works mostly consider external adversaries with a length-based fingerprint as the threat model.

Xiong \textit{et al.} formally define the level of risk of side-channel attack based on statistical features following a Differential Privacy algorithm \cite{diffprivacy}. Then, they show that the amount of padding in the packet packet corresponds directly to the level of obfuscation \cite{defending-side-channel}. One way to balance between privacy and performance is to adjust the level of padding according to link utilization.         

A number of solutions developed on SDN-based obfuscation systems \cite{adaptivepadding,extremesdn}. The former system dynamically adjusts the level of padding to reflect link utilization on a network \cite{adaptivepadding}. This system consists of middlebox devices (such as routers or gateways) that monitor utilization levels on their respective links, send this information to an SDN controller, and pad all packets at the link layer according to the padding level instructed by the SDN controller. In turn, the SDN controller adjusts the padding level based on the utilization information received from the middlebox devices and alerts them if there are any changes to the padding level. Uddin \textit{et al.} also propose an SDN system that provides a comprehensive set of defense mechanisms against fingerprinting, including traffic padding \cite{extremesdn}. The authors argue that packets from an IoT device in its initial setup phase are the most prone to fingerprints. Thus, they developed a flow policy for every IoT device from a given network to dictate the level of packet padding. Then, the assigned level of padding is adjusted depending on the device's life cycle. Packets from new devices are padded heavily to maximize the security at the price of bandwidth overhead. As the device ages, the padding level decreases as per the policy and reduces the bandwidth usage.

Pinheiro et al. \cite{nonadaptive} effectively pad packets at the link layer by choosing a valid random size for a packet. They use a local router to pad all inbound and outbound packets from a network. Specifically, the authors use the Random MTU approach; for each packet, the router picks a random size, \textit{R}, between the original packet size and its MTU, and pads the packet to \textit{R}. Furthermore, the packets are also tunneled with a Virtual Private Network (VPN) to provide further obfuscation. Packet padding can also be implemented within VPNs. For example, the authors in \cite{pingpong} design a custom VPN that pads all packets to the MTU, achieving complete obfuscation of packet-length information at the cost of substantial bandwidth overhead. While padding every packet to the MTU guarantees uniformity, the resulting overhead makes it impractical for many deployments. To address this, \cite{nonadaptive} proposes a randomized padding scheme that selects packet sizes uniformly at random within the allowed range, thereby reducing bandwidth consumption while still mitigating length-based fingerprinting attacks. However, this approach still leaks packet-length information since original packet sizes vary greatly. Furthermore, \cite{defending-side-channel} formally and empirically shows that the level of padding (bandwidth overhead) inversely correlates with the level of side-channel information leak. The authors from \cite{adaptivepadding} attempts to balance this trade-off by monitoring network load and dynamically adjusting the level of padding to reflect its environment using an SDN controller. Similarly, \cite{extremesdn} maintains padding levels specific to each device on the network and their life-cycle. 

The authors in \cite{kadron2022targeted} propose a targeted black-box side-channel mitigation approach called IoTPatch. It analyzes captured network traces, extracts size and timing features, quantifies information leakage per feature, and synthesizes a tunable packet padding and delaying mitigation strategy to minimize leakage and overhead based on user preferences. Uniform random noise is added to the packet sizes in \cite{alshehri2020} to defend against signature-based tunnel traffic analysis (STTA). The comparison of different packet padding schemes is presented in Table~\ref{table:padding-schemes}.

\begin{table}[]
\caption{Fingerprint prevention approaches based on packet padding.}
\label{table:padding-schemes}
\resizebox{\columnwidth}{!}{%
\begin{tabular}{|l|l|l|l|}
\hline
\textbf{Ref}  & \textbf{Padding Layer} & \textbf{Level of Padding}                                                   & \textbf{\begin{tabular}[c]{@{}l@{}}Privacy vs. Overhead\\ Priority\end{tabular}} \\ \hline
\cite{nonadaptive}       & Link                   & Random MTU                                                                  & Privacy                                                                         \\ \hline
\cite{app_level_shaping}     & Application            & Static                                                                      & Privacy                                                                         \\ \hline
\cite{defending-side-channel} & Link                   & \begin{tabular}[c]{@{}l@{}}Customizable based \\ on preference\end{tabular} & \begin{tabular}[c]{@{}l@{}}Privacy\\ and Overhead\end{tabular}                  \\ \hline
\cite{pingpong}     & Network                & Static                                                                      & Privacy                                                                         \\ \hline
\cite{adaptivepadding}     & Link                   & \begin{tabular}[c]{@{}l@{}}Adaptive to \\ Network Load\end{tabular}         & \begin{tabular}[c]{@{}l@{}}Privacy\\ and Overhead\end{tabular}                  \\ \hline
\cite{extremesdn}       & Link                   & \begin{tabular}[c]{@{}l@{}}Adaptive to\\ Device Lifecycle\end{tabular}      & \begin{tabular}[c]{@{}l@{}}Privacy\\ and Overhead\end{tabular}                  \\ \hline
\cite{kadron2022targeted} & Link & \begin{tabular}[c]{@{}l@{}}Customizable based \\ on preference\end{tabular} & \begin{tabular}[c]{@{}l@{}}Privacy\\ and Overhead\end{tabular} \\ \hline
\cite{alshehri2020} & Link & \begin{tabular}[c]{@{}l@{}}Uniform random noise\\ (15-40 bytes)\end{tabular} & Privacy \\ \hline
\end{tabular}%
}

\end{table}

\findingsbox{
Packet padding represents a fundamental building block for IoT privacy, yet current approaches only scratch the surface of its potential. The key insight that padding level directly correlates with privacy protection \cite{defending-side-channel} opens exciting avenues for intelligent, adaptive padding systems that can dynamically optimize the privacy-performance trade-off in real-time. Next-generation opportunities include: (1) AI-driven padding strategies that learn device-specific traffic patterns and adapt padding levels based on threat intelligence, (2) formal privacy guarantees through differential privacy frameworks that provide mathematically provable protection levels, and (3) cross-layer optimization where padding decisions coordinate with upper-layer protocols to minimize overhead while maximizing obfuscation. The transition from data driven evaluation to real-world deployment frameworks capable of injecting prevention mechanisms into actual device communications represents a critical breakthrough needed to validate effectiveness at scale. Looking forward, the critical challenge lies in developing lightweight padding algorithms that can run efficiently on resource-constrained IoT devices while providing formal privacy guarantees comparable to current gateway-based approaches.
}

%Most promisingly, the integration of customizable padding with emerging edge computing paradigms could enable privacy-preserving IoT ecosystems where protection adapts seamlessly to user preferences and environmental conditions.

\subsection{Traffic Injection} \label{sec: injection} 

Traffic injection refers to injecting artificially generated IoT traffic to hide actual user activity. As opposed to other fingerprint prevention methods, traffic injection does not strictly \textit{obfuscate} any traffic metadata; instead, it aims to decrease adversary confidence by overwhelming adversaries with a cloud of dummy traffic (e.g., user sessions). Pure traffic injection is also used in website fingerprinting prevention \cite{web_injection_defense}.

The work \cite{sniffmislead,rffingerprint,fats-origin, pgg_spa, MITRA} in this category differs in the approach of generating dummy traffic or user sessions. For example, SniffMislead \cite{sniffmislead} groups labeled IoT traffic based on semantic similarity, temporal variation, and frequency. Then, the profiled behaviors are used to create dummy sessions called \textit{phantom users} for injection. A similar user behavior-based traffic fingerprinting and injection is proposed in \cite{rffingerprint}, where the solution targets to defend against Fingerprint And Timing-based Snooping (FATS) \cite{fats-origin} attacks from local adversaries. Specifically, the labeled network traffic is grouped into intervals, and from each interval, the most consistently occurring activities are considered to generate the dummy activities in that interval in the future. The injection level is tunable in both the above approaches as there is a trade-off between the privacy and bandwidth overhead. The proposed system is deployed in a non-intrusive Raspberry Pi that acts as a gateway and injects dummy traffic over local WiFi.

Boukharrou \textit{et al.} apply traffic injection to limit inference on Smart Personal Assistants (SPAs) usage \cite{pgg_spa}. In particular, the authors build a \textit{Privacy Guard Gateway (PGG)} that mimics the functionality of real users, i.e., whenever they make an audio request to a SPA, the closely-placed PGG will capture that. Then, it shuffles its content using Natural Language Processing (NLP) to build a set of dummy requests and audibly repeat them to the real SPA. Thus, every time a user makes a request, the associated SPA also processes a set of dummy requests to obfuscate the genuine one. MITRA \cite{MITRA} is another traffic injection-based scheme that leverages different levels of obfuscation to mask network traffic without significant overhead. Table~\ref{table:injection_approaches} compares different traffic injection schemes.

\begin{table}[]
\caption{Fingerprint prevention approaches based on traffic injection.}
\label{table:injection_approaches}
\resizebox{\columnwidth}{!}{%
\begin{tabular}{|c|c|c|l|}
\hline
\multirow{2}{*}{\textbf{Ref}} & \multirow{2}{*}{\textbf{Injection Layer}} & \multirow{2}{*}{\textbf{Approach}}                                                                  & \multirow{2}{*}{\textbf{Tunability}}                                          \\
                              &                                          &                                                                                                     &                                                                                \\ \hline
\cite{sniffmislead}           & Network                                  & \begin{tabular}[c]{@{}c@{}}Generating\\ "Phantom Users"\end{tabular}                                & \begin{tabular}[c]{@{}l@{}}Tunable Adversary\\ Confidence\end{tabular}        \\ \hline
\cite{web_injection_defense} & Application                              & \begin{tabular}[c]{@{}c@{}}Injecting Dummy\\ HTTP Traffic\end{tabular}                              & \begin{tabular}[c]{@{}l@{}}Tunable Number\\ of Requests Injected\end{tabular} \\ \hline
\cite{park_injection}        & Network                                  & \begin{tabular}[c]{@{}c@{}}Injecting Activity\\ at Likely Intervals\\ Throughout a Day\end{tabular} & \begin{tabular}[c]{@{}l@{}}Tunable Number \\ of Covered Devices\end{tabular}  \\ \hline
\cite{pgg_spa}               & Physical                                 & \begin{tabular}[c]{@{}c@{}}Inject Audio\\ Requests\end{tabular}                                     & Not Tunable                                                                   \\ \hline
\cite{IoTGAN}                & Network                                  & \begin{tabular}[c]{@{}c@{}}Generating Adversarial\\ Traffic with GAN\end{tabular}                   & Not Tunable                                                                   \\ \hline
\cite{MITRA}                 & Network                                  & \begin{tabular}[c]{@{}c@{}}Injecting Dummy\\ Traffic\end{tabular}                                   & \begin{tabular}[c]{@{}l@{}}Tunable Level\\ of Obfuscation\end{tabular}        \\ \hline
\cite{DiffPrivInject}        & Network                                  & \begin{tabular}[c]{@{}c@{}}Injecting Decoy\\ Traffic\end{tabular}                                   & \begin{tabular}[c]{@{}l@{}}Tunable Balance\\ of Security and Privacy\end{tabular}  \\ \hline
\end{tabular}%
}

\end{table}

\findingsbox{
Traffic injection approaches show promise in reducing adversary confidence through behavioral obfuscation, yet fundamental questions remain about scalability and privacy implications of user profiling \cite{park_injection}. Key technical challenges requiring investigation: (1) formal metrics for injection effectiveness that quantify the relationship between dummy traffic volume and fingerprinting resistance, (2) privacy-preserving behavior modeling that generates realistic phantom users \cite{sniffmislead} without requiring detailed usage profiling, and (3) network-wide coordination protocols for distributed injection systems. The demonstrated non-intrusive deployment advantages \cite{sniffmislead} must be balanced against bandwidth overhead quantification and current work lacks systematic studies correlating injection levels with network performance metrics. Critical research gaps include: optimal injection timing strategies for interval-based approaches \cite{rffingerprint}, scalability analysis of gateway-based injection systems beyond single Raspberry Pi deployments, and robustness evaluation against adaptive adversaries who may distinguish between genuine and phantom traffic patterns. Additionally, cross-domain injection strategies that work across different IoT application areas (from smart assistants \cite{pgg_spa} to general device traffic) require systematic investigation to establish fundamental injection principles.
}

\subsection{Traffic Shaping} \label{sec: shaping}

Traffic rate metadata like transmission rate and inter-arrival time are unique among different IoT devices and their events. Thus, such metadata is commonly exploited by adversaries to infer sensitive information. Traffic shaping obfuscates this information by adopting a constant or randomly chosen traffic rate that stays uniform across different events. This approach has been popular against website fingerprinting and is proposed as one of the earliest defense techniques against general traffic analysis \cite{morphin2009, analysis_countermeasures, peekaboo2012}. 

In the case of constant rate-based shaping, one approach is called \textit{Independent Link Padding (ILP)}. It offers a constant transmission rate on a link independent of user activity \cite{spying, peekaboo2012}. Thus, the distribution of padded traffic rates gets strictly uniform to offer the highest level of obfuscation. Specifically, ILP maintains a flat transmission rate at the gateway device either by reducing the asking demand (rate) or adding dummy traffic to match the maintained rate. The first approach will introduce latency, whereas the latter demands extra bandwidth.

\textit{Stochastic Traffic Padding (STP)} \cite{smartshaping} also maintains a constant transmission rate for regular/genuine traffic, where the rate is higher than the regular traffic demand to avoid latency. However, when there is no genuine traffic, STP follows a stochastic distribution to decide when to transmit traffic at the exact chosen sending rate. Thus, instead of continuous dummy traffic transmission, STP follows a probabilistically chosen transmission time, which helps it drastically reduce the bandwidth demand compared to ILP. The mean bandwidth overhead can increase as the probability of dummy traffic transmission increases. However, the authors prove that the adversary confidence decreases exponentially with the linear increase in bandwidth overhead. Thus, although STP cannot perfectly obfuscate the traffic rate metadata, it can significantly decrease the level of the side-channel leak with a low bandwidth overhead. 
\textit{Dynamic Traffic Padding (DTP)} \cite{dtp} is a shaping method based on STP, and it improves upon STP by optimizing the tunable variables dynamically based on previous traffic. The authors mention that DTP is more computationally expensive than STP in terms of memory and storage. However, DTP offers a lower bandwidth overhead in comparison to STP due to finetuning variables. 

PrivacyGuard \cite{privacyguard} argues that STP fails to accurately model the correlation among different IoT events and users' behavior. Thus, it uses Generative Adversary Network (GAN) algorithm to generate more realistic dummy traffic and Long Short-Term Memory (LSTM) algorithm to accurately model the individual user behaviors. Similarly, the authors from \cite{fisher} use an unsupervised ML algorithm to model the correlation of IoT events to make the statistical distribution of packet timestamps more uniform. Furthermore, the authors of \cite{dp_shaper} propose an event-level differential privacy (DP) model to minimize latency. This DP-shaping model provides a general solution towards IoT traffic shaping irrespective of devices, by following a first-come-first-served queuing discipline to output traffic based on the input. On the flip side, the authors of \cite{truman_show} classify IoT devices into different types and suggest that different equipment (i.e binary state/multiple state) should apply different shaping methods. They further delve into the shaping scenario of a single device and of multiple devices in the same network. However, they do not suggest novel shaping approaches, nor do they elaborate on the overhead related to the different shapers (to best fit devices of different types).

All the above shaping systems can be implemented on a Raspberry Pi, acting as an IoT gateway. However, the authors of \cite{paros} point towards the necessity of additional hardware in shaping methods, and propose a novel reshaping system called PAROS. PAROS learns the traffic rate signature of the IoT device, after which it generates an artificial traffic signature based on hidden Markov Model. Based on this and partial padding, PAROS overcomes the necessity of more edge hardware and ensures user privacy, which is further reinforced through two real router deployments.

While the above approaches mainly shape traffic on the network layer, Datta \textit{et al.} build a library for developers to shape traffic on the application layer \cite{app_level_shaping}. Any nodes using this library can send and receive packets with uniformly distributed inter-arrival times and payload sizes. Another system that works in the application layer is Replacement AutoEncoder\cite{replace_autoenc}, which transforms sensitive parts of sensory data with less sensitive (but functionally similar) data. Here, the authors leverage a user-customized objective function for deep autoencoder to ensure user's behavioral privacy.
While network-level shaping implemented on a middlebox is only effective against external adversaries, application-level shaping is effective against both external and local adversaries since the IoT traffic is shaped before it leaves the device \cite{app_level_shaping}.

Alternatively, the authors of \cite{mac_shaping} consider data link device profiling attacks and propose a MAC layer traffic shaping technique where the injected dummy packets only exist between the WiFi link between the IoT device and its access point. This methodology boasts a zero bandwidth overhead, and prevents local adversaries from fingerprinting the IoT device.
Table \ref{table:shaping_approaches} compares the current traffic shaping schemes.

\findingsbox{
Current traffic shaping techniques demonstrate significant potential, but lack comprehensive evaluation across diverse IoT device types and network conditions. Critical research gaps include: (1) device-specific latency tolerance analysis to determine optimal shaping parameters for heterogeneous IoT environments, (2) formal verification of shaping indistinguishability against ML-based fingerprinting, particularly for techniques like PAROS \cite{paros} that claim statistical similarity to genuine traffic, and (3) multi-device coordination protocols for network-wide shaping that maintains temporal consistency across device interactions. The exponential relationship between bandwidth overhead and adversary confidence \cite{smartshaping} suggests opportunities for optimization frameworks that minimize overhead while maintaining protection guarantees. Emerging directions include hardware-accelerated shaping for resource-constrained devices, cross-protocol shaping strategies that work across WiFi, BLE, and cellular networks, and adaptive shaping policies that respond to real-time traffic analysis threats without compromising device functionality.
}

\begin{table}[]
\caption{Fingerprint prevention approaches based on traffic shaping. ILP = Independent Link Padding. DAE = Deep AutoEncoder. HMM = Hidden Markov Model}
\label{table:shaping_approaches}
\resizebox{\columnwidth}{!}{%
\begin{tabular}{|c|c|c|l|l|}
\hline
\multirow{2}{*}{\textbf{Ref}} & \multirow{2}{*}{\textbf{Shaping Layer}} & \multirow{2}{*}{\textbf{Approach}}                                      & \multirow{2}{*}{\textbf{Overhead}}                             & \multirow{2}{*}{\textbf{VPN?}} \\
                              &                                         &                                                                         &                                                                &                                \\ \hline
\cite{spying}                 & Network                                 & ILP                                                                     & \begin{tabular}[c]{@{}l@{}}Latency \&\\ Bandwidth\end{tabular} & Yes                            \\ \hline
\cite{trafficmorphing}        & Network                                 & ILP                                                                     & \begin{tabular}[c]{@{}l@{}}Latency \&\\ Bandwidth\end{tabular} & Yes                            \\ \hline
\cite{app_level_shaping}      & Application                             & \begin{tabular}[c]{@{}c@{}}Delaying \&\\ Padding packets\end{tabular}   & \begin{tabular}[c]{@{}l@{}}Latency \&\\ Bandwidth\end{tabular} & No                             \\ \hline
\cite{smartshaping}           & Network                                 & \begin{tabular}[c]{@{}c@{}}Randomized \&\\ Tunable Shaping\end{tabular} & Bandwidth                                                      & Yes                            \\ \hline
\cite{dtp}                    & Network                                 & \begin{tabular}[c]{@{}c@{}}Randomized \&\\ Optimized Shaping\end{tabular} & Memory                                                      & No                             \\ \hline
\cite{privacyguard}           & Network                                 & \begin{tabular}[c]{@{}c@{}}Randomized \&\\ Tunable Shaping\end{tabular} & Bandwidth                                                      & Yes                            \\ \hline
\cite{fisher}                 & Network                                 & Randomized                                                              & Bandwidth                                                      & No                             \\ \hline
\cite{dp_shaper}             & Network                                 & Differential Privacy                                                    & Bandwidth                                                      & No                             \\ \hline
\cite{replace_autoenc}        & Application                              & \begin{tabular}[c]{@{}c@{}}Transforming \\sensitive features \\using DAE\end{tabular} & Latency & No \\ \hline
\cite{mac_shaping}            & Data Link                                & \begin{tabular}[c]{@{}c@{}}Dummy packet \\injection\end{tabular} & Latency & No \\ \hline
\cite{paros}                  & Network                                  & \begin{tabular}[c]{@{}c@{}}Artificial signature \\generation using HMM \&\\partial padding\end{tabular} & Latency & No \\ \hline
\end{tabular}%
}

\end{table}

%\subsection{Hybrid Techniques}
%\textcolor{red}{can we name this section hybrid and blocking techniques and combine these last two subsections as there are not many papers under these categories? Also, please identify and exclude all GenAI papers and }

%Traffic padding, injection and shaping are the three main defending methods we summarized. 

\subsection{Hybrid and Other Techniques} \label{sec:hybrid}

This section presents the prevention schemes that combine most common prevention techniques (e.g., shaping, padding). Also, we outline techniques that block the IoT traffic to reduce the scope of fingerprinting. 

%There are fingerprint prevention schemes that combine the above approaches for obfuscation \cite{li2024homesentinel,10004987,vergutz2023data,barman2021every}, as shown in Table~\ref{table:merged_prevention}. These works usually build defense chains for real-world IoT environments rather than proposing novel methods.
%There are fingerprint prevention schemes combine the above approaches for obfuscation \cite{li2024homesentinel,  10.1145/3559613.3563191, vergutz2023data, barman2021every} as shown in Table \ref{table:merged_prevention}. \textcolor{red}{are you saying? these schemes mainly adopt existing prevention schemes instead of developing anew ones? They mainly focuses on the deployment feasibility of these schemes? --- Instead of providing a novel defending methodology, these works usually tend to build a defending chain for real IoT environment.}

\begin{table}[h]
\centering
\resizebox{\columnwidth}{!}{%
\begin{tabular}{|l|l|l|p{5cm}|}
\hline
\textbf{Ref} & \textbf{Layer} & \textbf{Approach} & \textbf{Main Contributions} \\ \hline
\cite{li2024homesentinel} & Application & Injection + Shaping & Obfuscation in mixed traffic environments (IoT and non-IoT devices). \\ \hline
\cite{10123697} & Application & Injection + Padding & Adaptive RL-based framework with fingerprintability assessment. \\ \hline
\cite{10004987} & Application & Injection + Shaping & Low-overhead defense against signature-based fingerprinting. \\ \hline
\cite{10.1145/3559613.3563191} & Application & Padding + Shaping & Evaluation under adaptive adversaries aware of obfuscation. \\ \hline
\cite{vergutz2023data} & Application & Padding + Shaping & Feature selection with PCA/FA and adaptive defense. \\ \hline
\cite{barman2021every} & Link & Injection + Shaping & BLE-specific limitations of standard fingerprint defenses. \\ \hline
\end{tabular}%
}
\caption{Hybrid fingerprint prevention approaches.}
\label{table:merged_prevention}
\end{table}

\subsubsection{Hybrid Techniques} 

IoTReGuard \cite{vergutz2023data} deploys principal component analysis (PCA) to choose the most effective features following factor analysis (FA) to group features that are correlated into fewer underlying factors. It then generates dummy traffic based on the chosen features following traffic padding \cite{adaptivepadding}. 
Similarly, the authors in \cite{10123697} propose an "adaptive" anti-fingerprinting system consists of two components. A one-class classifier as a "fingerprinting agent" to test given flows' probabilities being fingerprinted, then construct a reinforcement learning agent that tries various prevention actions including shaping and padding dynamically to adjust its prevention strategies in a real-time. 
Unlike the adaptive feature-based prevention schemes in \cite{vergutz2023data,10123697}, the authors in \cite{10004987} propose prevention methodology against fingerprinting specific features, especially signature-based fingerprinting, such as device events signature \cite{pingpong}. Instead of generating dummy network traffic in random distributions, authors generate the dummy network traffic in the time burst when device events likely occur. Also, traffic shaping is applied on these event traffic to obfuscate fingerprinting methods highly relying on device events' features.
%applying \cite{he2017adaptive} on the chosen features; then, apply padding scheme in \cite{adaptivepadding}.

The resiliency of various prevention schemes is analyzed in \cite{10.1145/3559613.3563191}. The authors tested eight different traffic padding and shaping schemes under the assumptions that attackers are aware of the presence of prevention systems without knowing further details (e.g., MTU padding or Gaussian padding). Surprisingly, they could successfully fingerprint both devices and their events using simple models like SVM and RF. Similarly, the resiliency of prevention schemes using BLE is tested in \cite{barman2021every}. The outcome from three tested schemes revealed that existing techniques are not a good fit in BLE environment. 
In particular, the static packet structure of BLE hinders the provisions for packet padding. Also, traffic shaping (e.g., extended communication duration) and traffic injection (e.g., higher traffic volume) would significantly increase communication overhead for the resource constrained BLE-based devices.

\subsubsection{Other Techniques}

Blocking some or all IoT traffic could be another approach to limit traffic analysis. The authors of \cite{closing_blinds} experiment to block all traffic from various commercially available devices, but find that most IoT devices are rendered completely useless without an internet connection. To retain device functionality, the authors of \cite{blocking}, \cite{xu2019privacy}, and \cite{yu2020iotremedy} block unnecessary traffic that has no effect on device functionality. Furthermore, \cite{voicetraffic} and \cite{leakypick} block traffic from voice assistants that users do not authorize. HomeSnitch \cite{homesnitch} and IoTSentinel \cite{iotsentinel} propose comprehensive defense systems that classify IoT events and allow users to block unwanted traffic. However, these partial blocking approaches are not designed to prevent traffic fingerprinting; thus, there is no evaluation showing their effectiveness for prevention.  

Private data transmission is not only a feature desired by consumers and researchers, but could also be a legal requirement. Panwar \textit{et al.} argue that there are emerging laws in California and the EU that tighten privacy requirements on data handling and transmission \cite{iotexpunge}. Thus, they develop a framework that automatically minimizes data storage duration in the cloud, consequently, reducing the attack surface for fingerprinting. Finally, Yao \textit{et al.} \cite{yao2019defending} fill a critical gap in the existing literature by incorporating user-specific privacy concerns into expert-driven privacy tool designs. Specifically, the design encompasses data transparency and control, security, safety, usability and user experience, system intelligence, and system modality.  
%\textcolor{red}{where is the comparison table?}

\findingsbox{
Hybrid prevention schemes reveal both the potential and challenges of multi-technique approaches, with limited systematic evaluation of technique interactions and cumulative effectiveness. Critical research needs include: (1) formal frameworks for technique composition that predict combined effectiveness and identify potential interference between padding, injection, and shaping, (2) adaptive adversary models that account for attackers aware of multiple prevention techniques \cite{10.1145/3559613.3563191}, and (3) resource optimization algorithms for technique selection based on device capabilities and threat models. The demonstrated limitations of existing techniques in constrained environments like BLE \cite{barman2021every} highlight the need for protocol-specific defense design. Future research directions include reinforcement learning frameworks for dynamic technique selection \cite{10123697}, energy-efficient hybrid architectures for battery-powered devices, and privacy-utility optimization that balances protection effectiveness with functional requirements across diverse IoT application domains.
}

\section{Generative AI in Smart Home Fingerprinting}
\label{sec:genai}

While previous sections covered a broad range of machine learning (ML) and non-ML methods for fingerprinting detection and prevention, recent advances in generative AI (GenAI) introduce fundamental paradigm shifts that transcend the limitations of traditional approaches. Unlike conventional methods that rely on discriminative classification of existing patterns or static obfuscation rules, GenAI enables temporal-aware synthesis, dynamic adaptation, and domain-conscious augmentation of network traffic. This represents a shift from reactive pattern recognition to proactive pattern generation, addressing core technical challenges in IoT fingerprinting: the cold-start problem for unseen devices, preservation of complex temporal dependencies, and real-time adaptation to evolving attack models.

\subsection{GenAI-Enabled Fingerprinting Detection}

Building upon conventional ML-based classifiers, GenAI addresses fundamental technical limitations through learned traffic synthesis rather than discriminative feature extraction. Traditional supervised classifiers fail catastrophically with unseen devices due to their dependency on labeled training data and inability to model complex temporal correlations. GenAI overcomes these constraints by learning underlying traffic generation processes, enabling synthesis of realistic patterns for devices never encountered during training.

IoTGemini~\cite{li2024iotgemini} exemplifies this paradigm shift through its Packet Sequence GAN (PS-GAN), which simultaneously preserves per-packet fidelity and long-term temporal dependencies, a capability impossible with conventional classification approaches. Unlike basic data augmentation that applies domain-agnostic transformations, PS-GAN learns IoT-specific device behavior profiles, enabling customized traffic generation that mimics real device operation. HSGAN-IoT~\cite{hsgan2024} introduces hierarchical semi-supervision to enable classifier training under minimal annotation, and FL4IoT~\cite{fl4iot2023} explores federated setups that preserve user privacy while improving generalization.

\subsection{GenAI-Enabled Prevention}

In contrast to fixed obfuscation strategies discussed in Section~\ref{sec:prevention}, GenAI-based defenses represent a fundamental departure from static obfuscation strategies by treating traffic perturbation as an adversarial learning problem. While traditional methods apply fixed rules (padding, timing manipulation) that attackers can characterize and circumvent, generative defenses continuously adapt their strategies to counter evolving fingerprinting models.

iPET~\cite{shenoi2023ipet} demonstrates this technical superiority through GAN-based tunable adversarial perturbations that allow users to specify exact bandwidth overhead constraints, a level of fine-grained control impossible with rule-based approaches. Critically, iPET introduces deliberate stochasticity in its GAN training process, preventing attackers from reverse-engineering the perturbation strategy. DiffPrivInject~\cite{DiffPrivInject} leverages differential privacy-constrained VAEs to generate perturbations that maintain compatibility with detection systems while preserving privacy guarantees. Feng and Sehatbakhsh~\cite{feng2025fingerprintingdevicesadversarial} employ stochastic diffusion models to generate decoy behaviors that require complete retraining of adversarial fingerprinting models.

HomeSentinel~\cite{li2024homesentinel} showcases end-to-end automation through its integrated GenAI pipeline: LightGBM automatically separates IoT traffic without manual labeling, while its Generative Adversarial Supervisory Network (GASN), a supervised variant of traditional GANs, generates dummy traffic that preserves IoT-specific metadata patterns. The system then applies GAN-learned traffic shaping by swapping time intervals between real and synthetic packets, creating realistic temporal patterns that fool fingerprinting classifiers.

Despite promising initial results, current GenAI approaches face several deployment barriers that represent key research directions. \textit{Computational constraints} remain critical: while iPET enables tunable overhead, comprehensive energy evaluations on battery-powered devices are lacking. \textit{Temporal consistency} presents challenges for long-term deployments, IoTGemini's PS-GAN excels at short sequences but maintaining behavioral coherence over days/weeks requires investigation. \textit{Adversarial adaptation} poses an arms race: as attackers adopt GenAI counter-strategies, defenses must evolve beyond current stochastic approaches. \textit{Scalability} remains largely unvalidated; evaluations typically involve <30 devices, while real smart homes contain hundreds of IoT endpoints. Finally, \textit{cross-domain generalization} is limited; models trained on smart home devices may fail against industrial or healthcare IoT with different behavioral patterns.

%Despite promising initial results, current GenAI approaches face several critical deployment barriers. Computational constraints remain a primary concern: while iPET enables tunable overhead, comprehensive energy evaluations on resource-constrained IoT devices are lacking. Temporal consistency presents challenges for long-term deployments; IoTGemini's PS-GAN excels at short sequences but maintaining behavioral coherence over extended periods (days/weeks) requires further investigation. The adversarial landscape poses an evolving arms race: as attackers adopt GenAI counter-strategies, defenses must evolve beyond current stochastic approaches toward provably robust mechanisms. Scalability validation remains limited, current evaluations typically involve fewer than 30 devices, while real smart homes contain hundreds of heterogeneous IoT endpoints. Cross-domain generalization presents another significant gap: models trained on smart home devices may fail when applied to industrial or healthcare IoT environments with fundamentally different behavioral patterns.

Key future research directions include: lightweight GenAI architectures achievable through model distillation and federated learning approaches, multi-scale temporal modeling that captures both packet-level timing and long-term usage patterns, provable adversarial robustness mechanisms against GenAI-powered attackers, universal IoT representation learning enabling effective cross-domain transfer, and comprehensive large-scale benchmarks incorporating thousands of devices with extended temporal coverage to validate real-world applicability.

\findingsbox{
GenAI approaches address fundamental limitations of discriminative classifiers through generative pattern synthesis, enabling classifier training for unseen devices and preserving temporal dependencies in traffic modeling. Unlike conventional augmentation, generative models learn device-specific operational semantics, producing training data that maintain IoT communication characteristics across encrypted traffic scenarios where supervised methods fail. Critical technical challenges remain: (1) computational efficiency; current GAN and VAE architectures require significant processing power incompatible with resource-constrained IoT devices, (2) temporal consistency validation; while PS-GAN preserves short-term dependencies, maintaining behavioral coherence over extended periods (days/weeks) lacks formal evaluation, and (3) domain transfer limitations; models trained on smart home traffic show poor generalization to industrial or healthcare IoT environments.
\\
GenAI-based defenses enable adaptive perturbation strategies through adversarial learning, offering tunable overhead constraints and stochastic perturbation models that resist reverse-engineering. However, fundamental research gaps persist: (1) scalability analysis; current evaluations \cite{li2024iotgemini, shenoi2023ipet} involve fewer than 30 devices while real deployments require hundreds of heterogeneous endpoints, (2) adversarial robustness quantification; formal analysis of defense sustainability against GenAI-powered attackers remains limited, and (3) energy consumption characterization; comprehensive battery impact evaluation in real IoT devices is lacking \cite{shenoi2023ipet}. Priority research directions include: lightweight model architectures through knowledge distillation, multi-scale temporal modeling frameworks, provable robustness bounds against adaptive adversaries, universal representation learning for cross-domain generalization, and large-scale benchmarks with extended temporal coverage for deployment validation.
}
% Prevention is the most mature so far

\section{Research challenges and future directions} \label{sec: challenge}

%Our comprehensive survey across 569 works reveals that IoT fingerprinting has matured along specialized detection pipelines, yet fundamental gaps persist that hinder real-world adoption, generalization, and resilience. These challenges extend beyond algorithmic refinements and require rethinking how fingerprinting methods are designed, evaluated, and deployed.

Our survey shows that IoT fingerprinting is no longer a niche proof-of-concept but a mature field, yet it remains hemmed in by structural blind spots. To move from lab prototypes to resilient, field-ready systems, we must confront these intertwined challenges:

\textbf{Imbalanced Research Focus and Evasion Vulnerability:} The 14:1 ratio between detection and prevention research (Section~\ref{sec:methodology}) underscores a systemic asymmetry that has enabled increasingly sophisticated fingerprinting models, including deep and generative approaches (Section~\ref{sec: discovering}), while progress in countermeasures remains limited. Most prevention schemes (Section~\ref{sec:prevention}) rely on static obfuscation strategies like burst shaping or padding, which are fragile under adaptive adversaries. Very few approaches incorporate adversarial training, and none offer formal robustness guarantees. Addressing this requires treating defense as a co-evolving, integrated component rather than a post hoc patch.

\textbf{Temporal Instability and Protocol Shifts:} Many surveyed methods degrade under firmware updates or device reboots, yet most datasets span only hours or days (Tables~\ref{tab:statistical_discovering}--\ref{tab:behavioral_policy}), failing to capture behavioral drift over time. Moreover, emerging standards like Matter and Thread introduce multi-hop routing, multicast discovery, and new stack behaviors that disrupt learned fingerprints. Existing techniques, often confined to TCP/IP or TLS layers (Figure~\ref{fig:iotstates}), lack abstraction across protocol generations. Research must investigate behavioral invariants that persist across stack evolution to support long-term resilience.

\textbf{Evaluation Misalignment and Deployment Blind Spots:} Of all works filtered  in Section~\ref{sec:methodology}, fewer than 15\% of them use public or longitudinal datasets, limiting reproducibility and ecological validity. Lab setups frequently omit deployment realities such as NAT traversal, encrypted traffic, or constrained hardware. While resource metrics like FLOPs or memory use are sometimes reported, latency impact and scalability under real-world load are rarely addressed. The absence of standardized, infrastructure-aware benchmarks hampers meaningful evaluation prior to deployment.

\textbf{Domain Fragmentation and Lack of Universal Features:} Most models are tailored to specific IoT verticals, smart homes, ICS, or healthcare systems (Section~\ref{sec: discovering}), and seldom evaluated for cross-domain transfer. Differences in traffic regularity, protocol usage, and regulatory constraints hinder generalization. Although flow-based and federated approaches (Section~\ref{sec: shaping}) show early signs of transferable representations, the field lacks a principled framework to discover and evaluate domain-invariant features.

\textbf{Adversarial Drift and Online Learning:} Despite growing awareness of evasion tactics like traffic mimicry or timing spoofing, most models remain static and offline-trained. Few support real-time model adaptation without retraining. Developing lightweight, continually learning models that maintain robustness under adversarial drift while avoiding catastrophic forgetting is a key open challenge.

\textbf{Toward IoT Foundation Models and Semantic Intelligence.}
Current methods remain bound to packet-level features and protocol-specific heuristics, missing the holistic, cross-layer context available in modern IoT (Sections~\ref{sec: discovering}, \ref{sec:prevention}). There is a clear opportunity to develop “foundation models” for IoT: self-supervised, multi-modal representations drawing from network traffic, RF traces, encrypted metadata, and device telemetry. Such models could power everything from zero-shot device discovery to privacy-preserving analytics, but demand rigorous solutions for federated learning, privacy compliance, and benchmarking at scale.

Tackling these challenges demands a new research paradigm, one that blends generative and discriminative techniques, unifies multi-layer data, and embeds security, privacy, and governance from the ground up. Only then will IoT fingerprinting mature into a truly deployable discipline.

\section{Conclusion} \label {sec: conclusion}

In this survey we have sistematically reviewed and compared the literature relative to fingerprinting IoT devices in smarthome environments.Our review shows that current research suffers from three systemic weaknesses. First, evaluation frameworks remain laboratory-bound, with most datasets spanning mere hours rather than the months or years required to validate real-world robustness. Second, the field lacks standardized benchmarks that account for deployment realities such as encrypted traffic, protocol evolution, and resource constraints. Third, prevention mechanisms rely on static obfuscation strategies that fail against adaptive adversaries.
The emergence of generative AI represents both opportunity and threat. While GenAI enables synthetic traffic generation and adaptive defenses, it equally empowers sophisticated attacks that render current prevention schemes obsolete. This technological arms race demands a fundamental shift from reactive to proactive security design.
Three critical research directions emerge. Domain-agnostic IoT foundation models could enable universal device representation and cross-vertical transfer learning. Adversarial robustness frameworks must replace current evaluation approaches that ignore co-evolving attack models. Large-scale longitudinal benchmarks spanning diverse IoT ecosystems are essential for validating deployment feasibility.
The field stands at an inflection point. Converting laboratory innovations into deployable security solutions requires embracing the adversarial nature of IoT fingerprinting rather than treating security as an afterthought. The billions of IoT devices entering homes worldwide demand this paradigm shift from academic exploration to robust, field-ready protection mechanisms.

\bibliographystyle{IEEEtran}
\bibliography{survey.bib}
\end{document}